\documentclass{emulateapj}
\usepackage{natbib}
\usepackage{longtable}
\usepackage{epsfig}
\usepackage{amssymb}

\newcommand{\kms}{km\,s$^{-1}$}

\newcommand{\be}{\begin{equation}}
\newcommand{\ee}{\end{equation}}
\def\cet{c_{\rm e2}}

\slugcomment{ApJ in press}

\shorttitle{SLACS VIII}
\shortauthors{Treu et al.\ }

\begin{document}

\title{The Sloan Lens ACS Survey. VIII. The relation between environment and internal structure of early-type galaxies
\altaffilmark{1}}

\author{Tommaso Treu\altaffilmark{2,}\altaffilmark{3}, Rapha\"el
Gavazzi\altaffilmark{2,}\altaffilmark{4}, Alexia Gorecki
\altaffilmark{2}, Philip J. Marshall\altaffilmark{2}, L\'{e}on
V. E. Koopmans\altaffilmark{5}, Adam S. Bolton\altaffilmark{6},
Leonidas A. Moustakas\altaffilmark{7}, and Scott Burles\altaffilmark{8}}

\altaffiltext{1}{ Based on observations made with the NASA/ESA Hubble
Space Telescope, obtained at the Space Telescope Science Institute,
which is operated by the Association of Universities for Research in
Astronomy, Inc., under NASA contract NAS 5-26555.  These observations
are associated with programs \#10174,\#10587, \#10886, \#10494,
\#10798.}
\altaffiltext{2}{Department of Physics, University of California,
Santa Barbara, CA 93106, USA ({\tt tt@physics.ucsb.edu, pjm@physics.ucsb.edu})}
\altaffiltext{3}{Sloan Fellow; Packard Fellow}
\altaffiltext{4}{Institut d'Astrophysique de Paris, UMR7095 CNRS - Universit\'e Paris 6, 98bis Bd Arago, 75014 Paris, France  ({\tt gavazzi@iap.fr})}
\altaffiltext{5}{Kapteyn Institute, P.O. Box 800, 9700AV Groningen,
The Netherlands ({\tt koopmans@astro.rug.nl})}
\altaffiltext{6}{Beatrice Watson Parrent Fellow,
Institute for Astronomy, University of Hawai`i,
2680 Woodlawn Dr., Honolulu, HI 96822 ({\tt bolton@ifa.hawaii.edu})}
\altaffiltext{7}{Jet Propulsion Laboratory, 4800 Oak Grove Dr, Caltech, MS169-327,
Pasadena CA 91109 ({\tt leonidas@jpl.nasa.gov})}
\altaffiltext{8}{Department of Physics and Kavli Institute for
Astrophysics and Space Research, Massachusetts Institute of
Technology, 77 Massachusetts Ave., Cambridge, MA 02139, USA ({\tt burles@mit.edu})} 

\begin{abstract}
We study the relation between the internal structure of early-type
galaxies and their environment using 70 strong gravitational lenses
from the Sloan ACS Lens Survey (SLACS). The Sloan Digital Sky Survey
(SDSS) database is used to determine two measures of overdensity of
galaxies around each lens: the projected number density of galaxies
inside the tenth nearest neighbor ($\Sigma_{10}$) and within a cone of
radius one $h^{-1}$ Mpc ($D_1$). Our main results are: 1) The average
overdensity is somewhat larger than unity, consistent with lenses
preferring overdense environments as expected for massive early-type
galaxies (12/70 lenses are in known groups/clusters). 2) The
distribution of overdensities is indistinguishable from that of
``twin'' non-lens galaxies selected from SDSS to have the same
redshift and stellar velocity dispersion $\sigma_*$. Thus, within our
errors, lens galaxies are an unbiased population, and the SLACS
results can be generalized to the overall population of early-type
galaxies. 3) Typical contributions from external mass distribution are
no more than a few per cent in local mass density, reaching 10-20\%
($\sim0.05-0.10$ external convergence) only in the most extreme
overdensities. 4) No significant correlation between overdensity and
slope of the mass density profile of the lens galaxies is found. 5)
Satellite galaxies (those with a more luminous companion) have
marginally steeper mass density profiles (as quantified by $f_{\rm
SIE}=\sigma_*/\sigma_{\rm SIE}=1.12\pm0.05$ vs $1.01\pm0.01$) and
smaller dynamically normalized mass enclosed within the Einstein
radius ($\Delta \log M_{\rm Ein}/M_{\rm dim}$ differs by
$-0.09\pm0.03$ dex) than central galaxies (those without). This result
suggests that tidal stripping may affect the mass structure of
early-type galaxies down to kpc scales probed by strong lensing, when
they fall into larger structures.
\end{abstract}


\section{Introduction}

The observed properties of galaxies correlate with their
environment. For example, the mix of morphological types depend on the
local number density of galaxies
\citep{Dre80a,Dre80b,P+G84,Dre++97,Tre++03,Pos++05,Smi++05,Cap++07,Cas++07} and dark
matter \citep{Lan++07}, with elliptical galaxies dominating in high
density regions. The star-formation rate and colors of galaxies also
scale with their local number density, or distance from the center of
clusters \citep[e.g.,][]{Bal++04,Got++03,Hog++04, Coo++08}.

The physical origin of these environmental trends has been studied
for over thirty years. A number of mechanisms have been proposed to
shut off or trigger star-formation as well as to modify the
mass-dynamical structure of galaxies. They include interactions with
other galaxies (mergers or harassment), with the dark-matter potential
of clusters and groups (tidal stripping; tidal compression,
harassment) and with the intracluster medium when present (starvation
or strangulation, ram pressure stripping). A review of the main
mechanisms, their effects on galaxies, and the range of environments
over which they operate is given by \cite{Tre++03}.  Although the
interaction between galaxies and their environment is not yet fully
understood, comprehensive analyses of large imaging and spectroscopic
datasets over a range of environments and cosmic times have shown that
a variety of mechanisms are at work, starting from the very outskirts
of clusters. Physical processes such as starvation and harassment start
to be effective at the group stage or when galaxies infall onto
clusters as groups, beyond the cluster virial radius. At higher
densities other mechanisms such as ram pressure stripping become
effective, further modifying the properties of galaxies.

To complement the progress based on traditional luminous tracers of
star-formation and morphology, it is clear that one can gain
additional insights by following the modifications to the
mass-dynamical structure of galaxies induced by the
environment. Empirical scaling laws connecting stellar populations
with kinematics, such as the fundamental plane and the Tully Fisher
relation, have suggested mild trends of the star-formation history of
early-type galaxies with environment, and that spiral galaxies are
dynamically perturbed as they enter massive clusters
\citep[e.g.][and references therein]{Mor++07c}.

Gravitational lensing provides an additional tool to address the
connection between galaxies and their environment. By measuring total
mass directly --- rather than through optical tracers ---
gravitational lensing gives a direct handle on the transformations of
the mass dynamical structure. For example, weak galaxy-galaxy lensing
studies in clusters have shown that dark-matter halos of infalling
galaxies are tidally truncated \citep{NKS02,Gav++04,Nat++07}.

In this paper we exploit the large and homogeneous sample of strong
gravitational lenses discovered by the Sloan Lenses ACS Survey
\citep[][hereafter collectively SLACS, or SLACS Papers I through
VII]{Bol++06, Tre++06, Koo++06, Gav++07, Bol++08a, Gav++08, Bol++08b}
to study the connection between environment and mass-dynamical
structure at the typical scale of galaxy strong lensing
(i.e. $\sim10$kpc). At these scales the mass distribution is more
directly connected with morphology and stellar populations than at the
larger scales probed by weak lensing.

Two main effects are expected. On the one hand we expect that tidal
truncation by an external potential could steepen the local mass
density slope, as suggested for the case of PG1115+080
\citep{Imp++98,T+K02b}, by numerical simulations \citep{DKF07}, and by
an analysis of the first 15 lenses in the SLACS sample \citep{Aug08}.
If this is the case, then we would expect empirical scaling relations
---- such as the correlation between velocity dispersion of the best
fit lens model and stellar velocity dispersion and the mass plane
\citep[MP;][]{Bol++07}--- to depend on the environment. On the other hand,
since lensing only measures projected mass, a high density environment
with a relatively smooth and shallow embedding dark-matter halo could
mimic a shallower local slope and therefore skew measurements in the
opposite direction.  One of the goals of this paper is to clarify
these issues in order to improve our understanding of the internal
structure of lens galaxies, and therefore of early-type galaxies in
general if lenses are an unbiased subset.

In addition to providing a diagnostic of the interaction between
galaxies and their environment -- and of the internal structure of
early-type galaxies -- this study also has repercussions for a number
of applications of gravitational lensing.  For example, the degeneracy
between local mass density slope and external mass density is the
dominant source of error in measuring the Hubble Constant from
gravitational time delays
\citep[e.g.,][]{Koc02,Koo++03,Mou++07,Ogu07a,Suy++08}. Discovering trends
with the local environment may help reduce systematic errors in these
measurements. Similarly, the effects of the local environment need to
be taken into account to interpret weak galaxy-galaxy lensing results,
as well as to do precision cosmography based on lensing statistics.

In previous Papers (I,II,IV,V) we showed that the SLACS lenses are
statistically indistinguishable within the current level of
measurement errors from control samples in terms of properties such as
size, luminosity, surface brightness, location on the Fundamental
Plane, and weak lensing signal, and thus our results could be
generalized to the overall population of early-type galaxies. In this
paper, we address the question of whether the environment of SLACS
lens galaxies differs from that of non-lens galaxies with the same
properties, using samples of ``twin galaxies'' (or simply ``twins'')
selected to have the same stellar velocity dispersion and redshift.

The paper is organized as follows. In Section~\ref{sec:data} we
briefly review the SLACS sample, describe the selection of a sample of
twin galaxies from the Sloan Digital Sky Survey (SDSS), and present
our new measurements of local environment.  In
Section~\ref{sec:comparison} we discuss whether the environments of
lens galaxies are special compared to those of non-lens galaxies with
similar properties. In Section~\ref{sec:intstuc} we explore the
dependence of the internal structure of early-type galaxies on the
environment.  In Section~\ref{sec:disc} we discuss our findings, and
in Section~\ref{sec:sum} we provide a summary.

Throughout this paper magnitudes are given in the AB scale. We assume
a concordance cosmology with matter and dark energy density
$\Omega_m=0.3$, $\Omega_{\Lambda}=0.7$, and Hubble constant
H$_0$=100$h$kms$^{-1}$Mpc$^{-1}$, with $h=0.7$ when necessary.

\section{Sample Selection and Data Analysis}

\label{sec:data}

Section~\ref{ssec:samp} summarizes the SLACS selection process,
reviews the properties of the SLACS lenses analyzed in this study, and
describes the selection of a control sample of ``twins'' from the
Sloan database. Section~\ref{ssec:meas} presents our measurements of
environment.

\subsection{SLACS lenses and SDSS twins}
\label{ssec:samp}

The sample analyzed in this paper is composed of the early-type lens
galaxies identified by the SLACS Survey. We include all the 70 lenses
classified as definite (grade ``A''), including 63 successfully
modeled as a single singular isothermal ellipsoid. The seven unmodeled
lenses (see paper V for details) include three classified as complex,
i.e. where a single singular isothermal ellipsoid is not a good
description due to the presence of a very close nearby companion. We
do not exclude those from the sample not to introduce biases against
high density environments. The vast majority of the lenses are
morphologically classified (paper V) as early-type galaxies (62/70), a
small fraction are classified as disk galaxies (6/70), and 2/70 have
ambiguous morphology.

A full description of the SLACS Survey and the selection process ---
together with images of all the lenses discovered with the Advanced
Camera for Surveys (ACS) --- is given in papers I and V of this series
(Bolton et al.\ 2006, 2008; see also Bolton et al.\ 2004, 2005 and the
SLACS website at www.slacs.org). For easy reference, we give here a
brief summary. First, lens candidates are found in the SDSS database
by identifying composite spectra made of a quiescent stellar
population and multiple emission lines at a higher redshift. The
spectra are taken from the Luminous Red Galaxies sample (Eisenstein et
al.\ 2001) and the MAIN galaxy sample (i.e. galaxies with $r$-band
Petrosian magnitudes brighter than $17.77$ and $r$-band petrosian
half-light surface brightness brighter than $24.5$ mag arcsec$^{-2}$;
Strauss et al.\ 2002). Quiescent spectra are selected from the MAIN
sample by imposing a limit on the rest-frame equivalent width of
H$\alpha$ $<4$\AA.

Second, the probability of the candidate being a lens as opposed to a
chance overlap within the fiber is computed based on the SDSS stellar
velocity dispersion, the lens and source redshifts, and an isothermal
mass model.  In this model, the probability of being a lens is a
monotonically increasing function of Einstein radius and therefore,
for any given source and lens redshift, of velocity dispersion.  As
discussed in Papers II and V, velocity dispersion and redshift appear
to be the two most important parameters characterizing the properties
of the lenses. 

Third, the most promising candidates have been imaged with ACS on
board the {\em Hubble Space Telescope} (HST) to confirm the lens
hypothesis (Programs 10174, 10587, 10886). Confirmed lenses are
followed-up with ACS and NICMOS to obtain deep three color images
(Programs 10494, 10798, 11202). After the failure of ACS, Programs
10886, 10494 and 10798 have been switched to the Wide Field and
Planetary Camera 2 (WFPC2). The follow-up programs are still ongoing.
The present analysis is based on the ACS dataset described in paper V.

For each lens, the following quantities are available from Paper V:
coordinates, redshift of the lens and source, F814W magnitude,
rest-frame V-band magnitude, effective radius, axis ratio, position
angle, morphology, stellar velocity dispersion, and Einstein radius
(as derived from a singular isothermal ellipsoid lens model).

In order to compare the environment of SLACS lenses with those of
non-lens galaxies, we randomly select for each SLACS lens a sample of
100 ``twins,'' i.e. galaxies with virtually the same velocity
dispersion and redshift.  As discussed in papers I, II and V, SLACS
lenses are not in general representative of the full parent sample,
but appear to be effectively velocity dispersion selected. Therefore,
to further investigate any selection bias due to lensing it is
important to compare to a similarly velocity dispersion selected
sample of twins.

Note that the SLACS series adopts stellar velocity dispersion
measurements based on the specBS pipeline developed by David Schlegel,
with cuts in signal-to-noise ratio
\citep[see discussion in][]{Bol++08a}. This estimate is not available
for a few of the lens galaxies. Therefore, in order to ensure uniform
and self consistent measurements for all lenses, we use the standard
stellar velocity dispersion from the SDSS-DR6 database \citep{Ade++08}
as a matching parameter between lenses and twins. 

One hundred twins are found for each lens. Fifty twins are chosen to
have velocity dispersion just above that of the lens and the remaining
fifty have velocity dispersions just below that of the lens. This
ensures an unbiased comparison sample (cf.\ discussion in papers II
and V). The number of twins per lens is sufficiently high to make
errors on the average properties of the comparison sample negligible
with respect to those on the SLACS lenses. At the same time, one
hundred twins are available for every lens and they are a tractable
number for quality control, as described in~\S~\ref{ssec:meas}.

\subsection{Environment measures}
\label{ssec:meas}

We measure two standard environment parameters \citep[see,
e.g.,][]{Coo++05} using the SDSS database: i) the projected number
density of galaxies inside the circle of radius equal to the projected
distance to the tenth nearest neighbor of the lens $\Sigma_{10}$
\citep[see][]{Dre80b}; ii) the projected number density of galaxies
within a 1 $h^{-1}$ Mpc radius circle $D_1$ centered on the
lens\footnote{The arbitrary scale is chosen to represent the typical
scale of a cluster.}. For brevity we refer to those two measures of
environment as ``local'' and ``global'' density. We choose the term
local and global to emphasize that $\Sigma_{10}$ is an adaptive
measure that can be sensitive to overdensities on small scales, like
subclumps or cores of clusters of galaxies, while $D_1$ is by
construction sensitive to overdensities on the typical scale of an
entire cluster. Therefore in general we expect the dynamic range of
$\Sigma_{10}$ to be larger than that of $D_1$. Both quantities are
measured in units of Mpc$^{-2}$.

In practice we proceed as follows. For each lens we query the SDSS
database and retrieve a catalog (hereafter ``cone'') of all the
objects within a one $h^{-1}$ Mpc radius. In the present analysis we
limit ourselves to the photometric catalogs, since spectroscopic
catalogs are rather sparse (except for the lowest redshift lenses) and
are biased by the fiber positioning strategy of the SDSS. We then
select objects belonging to the environment of the lens based on their
magnitude and photometric redshifts \citep{Csa++03}. We adopt the
following magnitude limits and redshift intervals: $i'<i'_{\rm
lens}+3$, $z_{\rm lens}+\delta z_{\rm lens}>z>z_{\rm lens}-\delta
z_{\rm lens}$, with $\delta z_{\rm lens}=0.03 (1+z_{\rm lens})$. For
brevity, galaxies within these magnitude and photo-z intervals are
referred to as ``neighbors.''  The magnitude limit ensures that even
the faintest galaxies in our environment samples are brighter than
$i'=21$, ensuring accurate photometric redshifts \citep{Csa++03},
while the redshift range represents a compromise between completeness
and purity. In general, photo-z slices much smaller than our choice
have to be discarded because they become smaller than photo-z accuracy
and comparable with peculiar velocities expected in clusters, while
photo-z slices much larger than our choice tend to dilute the
signature of the local environment. We have verified that our
measurements are robust with respect to reasonable changes of
magnitude cuts and photometric redshift slices.  In order to prevent
biases due to edge effects or gaps in the photo-z catalog (due, e.g.,
to bright stars) all catalogs are visually inspected. Four of the
lenses have incomplete cones, leaving a total sample of sixty-six
lenses with reliable environment measures\footnote{For six of the
sixty-six lenses there are fewer than 10 neighbors within each cone
(at least 7 in any case). In those cases the local density is set
equal to the global density. Thus $D_1\leq\Sigma_{10}$, by
definition.}.

We note that the absolute values of $\Sigma_{10}$ and $D_1$ depend on
the adopted magnitude and redshift limits
\citep[e.g.][]{Tre++03}. Therefore, in order to interpret our
environment measures we need to consider properly normalized
quantities. To this aim, we perform the same measurements around
pointings randomly selected from the SDSS archive. The average surface
density of neighbors measured from one hundred random fields with the
appropriate cuts in magnitude and redshift for each lens are referred
to as $\langle \Sigma_{10} \rangle$ and $\langle D_{1}
\rangle$, respectively for the two indicators. Note that 
the average density of the Universe is independent of the adopted
aperture, so the two quantities are the same when averaged over random
pointings. The uncertainty on the mean is also very similar, since a
random cone of radius one h$^{-1}$ Mpc contains approximately 10
galaxies for the typical magnitude and redshift cuts of our survey.
Throughout this study we will adopt as environment measures local and
global {\it overdensities} defined as $\Sigma_{10}/\langle
\Sigma_{10} \rangle$ and $D_1/\langle D_{1} \rangle$, respectively. 
As discussed, e.g., by \citet{Hog++03} and \citet{Coo++06b} this
procedure allows one to compare data of different intrinsic
depth. Furthermore, the distribution of density for the random fields
allows us to estimate the uncertainty due to sample variance,
photometric redshift uncertainties, and incompleteness. To be
conservative, we adopt as errors on the overdensity the sum in
quadrature of the Poisson uncertainty on the density measures, and of
the sample variance of the random fields as an upper limit to all
other uncertainties.

To address the issue of whether the environments of lenses are
special, we perform the same measurements for the twins of each
lens. After visual inspection to reject a few corrupted catalogs by
bright stars or survey edges, we compute $\Sigma_{10}$ and $D_1$ for
the twins. Table~\ref{tab:env} lists for each lens $\Sigma_{10}$ and
$D_1$, as well as the average of the same quantities measured for the
random fields ($\langle \Sigma_{10}
\rangle=\langle D_{1} \rangle$), and for the twins ($\langle
\Sigma_{10} \rangle_t$ and $\langle D_{1} \rangle_t$).

In addition, we compute the number of objects in each cone
significantly brighter (i.e. $i'<i'_{\rm lens}-1$) than the lens to
determine whether the galaxy is the central galaxy of the ``cone''
(and thus presumably of the halo associated with the overdensity), or
rather a satellite to a larger galaxy. The difference in magnitude is
chosen as a compromise to identify significantly more massive
companions while preserving enough statistics. Our conclusions are not
changed significantly if a threshold of 0.5 or 1.5 magnitudes is
adopted instead.

We define as ``satellite'' galaxies those with at least one neighbor
brighter than the lens galaxy by at least one magnitude, and
``central'' galaxies all the others. In the remainder of this paper we
will compare the properties of satellite galaxies with those of
central galaxies to assess whether this affects their internal
properties \citep[see also][]{Aug08}. The connection with the mass
profile at larger radii, as determined from a joint weak and strong
lensing analysis (paper IV), will be examined in future work.

\begin{figure}
\begin{center}
\resizebox{\columnwidth}{!}{\includegraphics{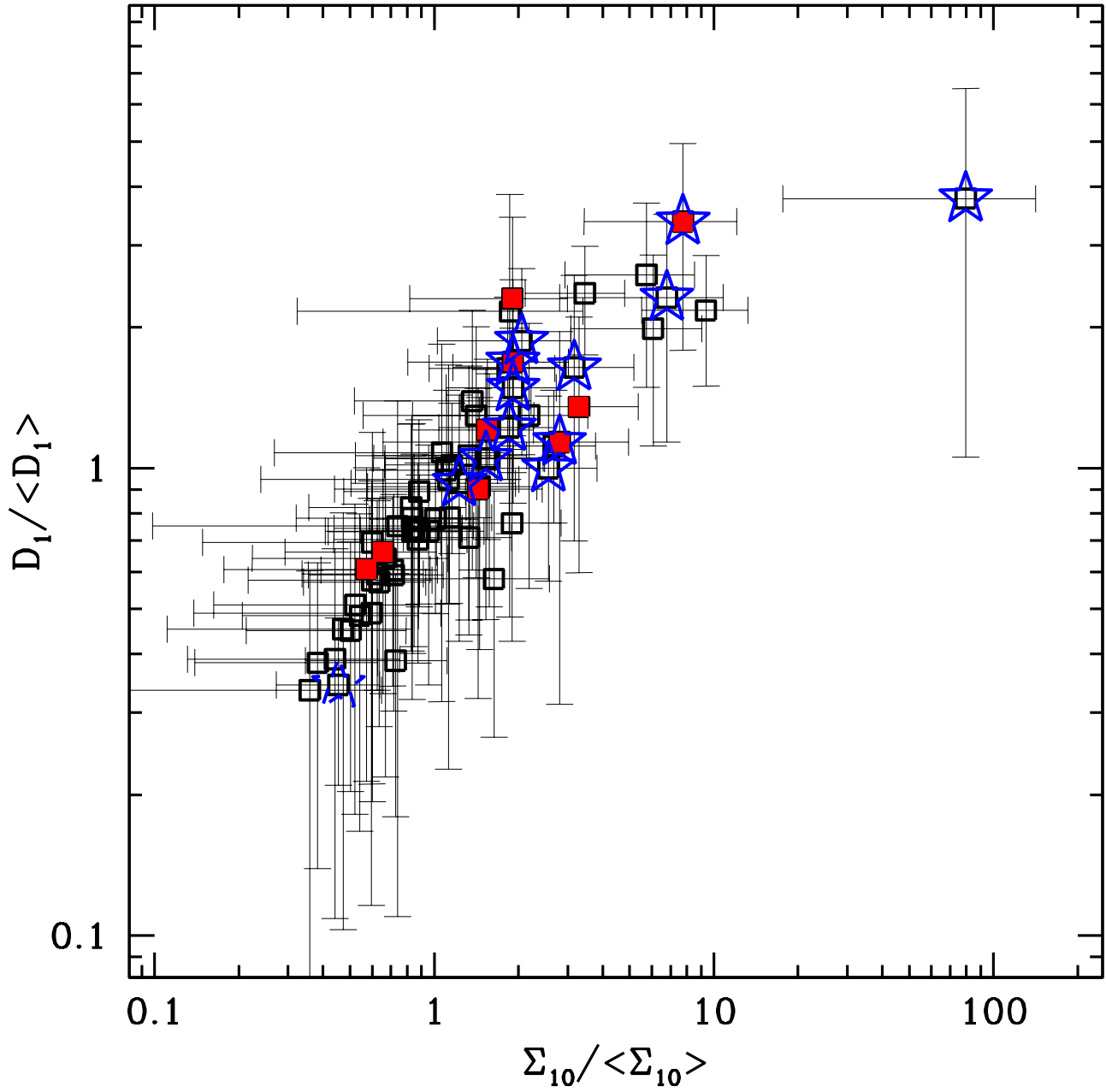}}
\end{center}
\figcaption{Distribution of environment estimators for the SLACS
sample.  Matches with known groups/clusters are identified by blue
open stars. Red solid points identify ``satellite'' galaxies, as
described in the main text. The lens with the highest overdensity
(J1143-0144) is known to be part of the optically selected cluster
Abell~1364 (Figure~\ref{fig:1143}). The lens associated with a known
group/cluster at the lower end of the overdensity range (J0252+0039;
identified by a dashed blue star), is associated with cluster
candidate NSCSJ025225+003540. Inspection of the field via SDSS
multicolor images shows no sign of a cluster near. We thus consider
this identification as spurious (see text for details).
\label{fig:densities}}
\end{figure}

Figure~\ref{fig:densities} shows local overdensities versus global
overdensities for the sample of lenses. Some of the lenses live in
considerable overdensities, as may be expected for massive early-type
galaxies.

As a final check, we used the NASA Extragalactic Database to look for
known clusters in the vicinity of the lenses. We restricted our search
to known clusters within $10$ arcminutes, which is 20 times the
typical Einstein radius of a massive cluster at these redshifts. We
expect that the shear and convergence contributed by massive clusters
more distant on the sky than 10 arcminutes will be less than 2.5
percent at the location of the lens, and even less for groups,
assuming conservatively an isothermal profile.  For comparison, this
10 arcminutes scale corresponds to approximately 2\,Mpc at the typical
redshift of the SLACS lenses ($z\approx0.2$), i.e. comparable to the
virial radius of a large cluster where the shear is only a few percent
\citep[e.g.,][]{Kne++03}. Therefore, we can
safely neglect the more distant clusters in the present analysis.

Table~\ref{tab:env} lists for each match the name of the cluster, the
distance on the sky in arc-minutes and the redshift of the cluster (if
known). In case of multiple matches for one lens (frequently because
the same cluster is identified by more than one survey), we list the
clusters in increasing order of three-dimensional distance from the
lens. Clusters with no known redshift are listed last. We consider a
cluster physically associated to a lens if the redshifts match to
better than $0.02$, where the interval is chosen to take into account
the typical errors in photo-z of SDSS based cluster searches
\citep{Koe++07}. In support of the reliability of our environment
measures, 12 of the 13 lenses associated with known clusters do lie in
overdense regions. The highest overdensity is associated with lens
J1143-0144 and the known cluster Abell~1364 (also identified by the C4
and MaxBCG surveys); a SDSS image of the field is shown in
Figure~2. The only exception is the lens J0252+0039, which is $6.6'$
from candidate group/cluster NSCSJ025225+003540 (estimated redshift
$z=0.27$). The candidate cluster is only detected by one of the
methods applied by
\citet{Lop++04} (Voronoi tesselation) and is not visible in the SDSS
image of the field.  We therefore consider this group/cluster
identification as spurious and ignore it in the rest of the
paper. Visual inspection of the SDSS images of the other 12 matches
confirms their identification as groups/clusters.

\section{Are the environments of lenses special?}

\label{sec:comparison}

To quantify whether the environment of lenses are special, we compare
in Figure~\ref{fig:plotlogs2} the distribution of normalized global
and local galaxy density for the lenses and for the twins. To
facilitate the comparison, the histogram for the twins has been
renormalized to the number of lenses.  Lens galaxies tend to live in
overdense environments -- as indicated by the average values of
$D_1/\langle D_1 \rangle$ and $\Sigma_{10}/ \langle \Sigma_{10}
\rangle$ somewhat larger than unity -- although the range is
broad. This is expected due to the clustering of massive early-type
galaxies
\citep[e.g.,][]{D+G76,Hog++03,Coi++08,Men++08}. However, the same
holds for the twins. The distribution of density measures appears to
be indistinguishable for the two samples. The Kolmogorov-Smirnov
statistic may suggest a marginal level of difference (the probability
that $\Sigma_{10}$ and $D_1$ of the lenses and twins are drawn from
the same distribution is $\sim 8-9$\%). However, the average values do
not seem to corroborate any difference: the mean local overdensity is
somewhat larger for lenses than for twins (less than 68\%CL), while
the opposite is true for the global overdensity (at less than 95\%CL).
The distributions are skewed, with the average larger than the median
due to the extended tail towards high densities.  Therefore it is
useful to compare the median and semi-quartile intervals, which are
less sensitive to the tails of the distribution.  The median
(semi-interquartile intervals) of $\Sigma_{10}/\langle
\Sigma_{10}\rangle$ is 1.15 (0.59) for the lenses and 1.40 (0.94) for
the twins. The median (semi-interquartile intervals) of $D_{1}/\langle
D_{1}\rangle$ is 0.90 (0.34) for the lenses and 1.03 (0.42) for the
twins.

Thus, we conclude that -- within the current measurement errors of our
survey -- lens galaxies live in the same environment as similar
non-lens galaxies.  Note that in principle our lenses are selected
also on the basis of magnitude and colors, at least for the luminous
red galaxies subset. However, our finding extends the conclusion of
Papers II and V that the dominant parameter is velocity dispersion:
once velocity dispersion is matched, lenses are indistinguishable from
``twins.'' This is consistent with the tight correlation between
velocity dispersion and global parameters of early-type galaxies
\citep[e.g.,][]{BBF92}. In conclusion, this result corroborates our
previous findings that SLACS lens galaxies are an unbiased sample of
massive early-type galaxies, and our working assumption that our
lens-based results can be generalized to the overall population.

\section{Environment and internal structure of early-type galaxies}

\label{sec:intstuc}

In this section, we investigate the dependence of the internal
structure of lens galaxies on the environment. We first look in
\S~\ref{ssec:PA} at the mis-alignment between mass and light, which we
expect to increase with increasing local and global density as a
result of the external potential associated with the large scale
structure. In Section~\ref{ssec:ff} we analyze the dependence of the
slope of the mass density profile of the lens galaxy on environment,
to probe tidal stripping of dark-matter halos.  In
Section~\ref{ssec:ML} we study the run of projected mass with radius
and the mass plane scaling law, in relation to the environment, as a
probe of variations of internal structure and of extra mass associated
with overdensities.

\subsection{Alignment of mass and light}
\label{ssec:PA}

In this section we compare the position angle of the best fit singular
isothermal ellipsoid model with no external shear to the position
angle of the light distribution. Because of the significance of the
stellar contribution at the scales probed by the Einstein radii of
SLACS lenses, we expect that the two will be aligned if there is no
significant external potential. Vice-versa, the r.m.s. amplitude of
the misalignment is a measure of the external perturbing potential
\citep{KKS97,Koo++06}. In Paper III we used the small
r.m.s. amplitude of the misalignment to show that the average external
shear must be smaller than 0.035 (see also Paper VII).  In this
Section we investigate trends with the environment.

Figures~\ref{fig:S10DPA} and~\ref{fig:D1DPA} show the offset between
the position angle of the light and that of the mass distribution as a
function of the local and global overdensity. As discussed in Paper
VII, the position angle can only be well measured if the light and
mass distributions are not circular and the lens model provides
sufficient constraints. Therefore, only the 25 objects with
significant ellipticity (axis ratio $<0.9$) and with well measured
mass PA (the ``ring subset'') are shown.

The correlation between alignment and local and global environment is
significant.  Fitting for example a power law relation, we find
$\sin^2(\Delta {\rm PA})=(0.14\pm0.05)\log(D_1/\langle D_1
\rangle)+0.03\pm0.01$ and $\sin^2(\Delta {\rm
PA})=(0.10\pm0.03)\log(\Sigma_{10}/\langle\Sigma_{10}\rangle)+0.02\pm0.01$.

We do not expect a simple relation between the two quantities, as a
galaxy sitting right at the center of a cluster could in principle
experience no external shear. However, in general, lenses in overdense
environments will be more likely to be affected by nearby companions,
causing a misalignment between the position angle of the stellar mass
and that of the total gravitational potential, since the external
perturber is presumably located in a random direction. Using
Equation~22 from \cite{KKS97} we estimate that for the mean mass axis
ratio of our sample $(b/a)_{\rm SIE}\approx0.77$, external shear of
order 0.05-0.06 is needed to match the average observed $\sin^2 \Delta
{\rm PA}\sim0.1$ in the overdense regions, while external shear
appears to be negligible in the underdense regions.  Assuming that the
external perturbers can be approximated to first order as singular
isothermal spheres, this would mean that -- even for the larger
overdensities -- the external contribution to the local surface mass
density (convergence) is of order $\lesssim$10 \% within the Einstein
radius of the main lens.

\subsection{Mass density profile}
\label{ssec:ff}

Figures~\ref{fig:S10ff1} and~\ref{fig:D1ff1} show f$_{\rm SIE}$ -- the
ratio between the central stellar velocity dispersion $\sigma_*$ and
that of the best fitting singular isothermal ellipsoid $\sigma_{\rm
SIE}$ -- as a function of the local and global overdensity. This ratio
is a direct empirical measure of the local density slope, expected to
be close to unity for isothermal profiles, larger than unity for
steeper slopes and smaller than unity for flatter slopes
\citep[e.g.,][]{T+K02a,T+K04}.  We thus use this quantity to study the
effect of the environment on the local mass density slope of galaxies
at kpc scales, since this is an excellent proxy for the total mass
density slope $\gamma'$ (such that $\rho_{\rm tot}\propto
r^{-\gamma'}$) 
measured from a full lensing and dynamical model as described by
\citet{Koo++06}. Error bars on $\gamma'$ are derived from the
posterior probability distribution function on the parameter as
described by \citet{Koo++06}.  The full results of lensing and
dynamical models will be presented in a forthcoming paper (Koopmans et
al.\ 2008, in prep). Figure~\ref{fig:fgamma} shows the goodness of the
proxy for the 58 objects in common between this study and Koopmans et
al.\ (2008): $\gamma'-2=(1.99\pm0.07)(f_{\rm SIE}-1)+(0.006\pm0.008)$
(see Figure~\ref{fig:fgamma}), for isotropic models. It is important
to emphasize that $\sigma_*$ and $\sigma_{\rm SIE}$ are the input to
determining $\gamma'$ via a well defined set of equations
\citep{Koo05}. Therefore, the errors are highly correlated along the
direction of the best fit relation $\delta \gamma'/\gamma' \sim \delta
f_{\rm SIE}/ f_{\rm SIE}$ and the range covered by both quantities in
Figure~\ref{fig:fgamma} is a representation of the intrinsic scatter
as well as the observational error. The small residual scatter in the
transformation between $\gamma'$ and $f_{\rm SIE}$ is due to the range
of Einstein radius to effective radius ratios spanned by the SLACS
sample. Note also that this best fit transformation is obtained for
isotropic models with Hernquist stellar profiles, and we expect it to
be different for other types of models.  Also, we would expect to find
a different transformation when the average ratio between effective
radius and Einstein radius is significantly different, e.g. for the
LSD sample \citep{T+K04}. A comprehensive analysis of $\gamma'$, its
statistical distribution and dependence on orbital parameters, as well
as host galaxy properties, will be given in Koopmans et al.\ (2008, in
prep).

The plots show two basic facts. First, the correlation between local
and global and environment and f$_{\rm SIE}$ is not
significant. Fitting a linear relation between f$_{\rm SIE}$ and the
logarithm of the local and global overdensities gives slopes of
$0.02\pm0.03$ and $0.10\pm0.06$ respectively. If it were present, a
correlation would indicate that mass density profiles tend to be
steeper in overdense environment perhaps as a result of tidal
truncation of the galactic dark-matter halos by external fields
\citep[as suggested by some numerical simulations,
e.g.,][]{DKF07}. The lack of correlations suggests that either the
effect is not strong enough to be detected at the kpc scales probed by
the SLACS Einstein Radii given the uncertainties in both the local
slope and the environment measures, or that the contribution of the
environment to the projected mass density is sufficient to
counterbalance the effect.  This latter explanation, however, does not
appear to be supported by the data, as the local density slope of the
lenses belonging to known clusters is indistinguishable from that of
the ``isolated'' central lenses (Figure~\ref{fig:histonbobjff1}).

Second, our result hints that the mass density profiles of
``satellite'' galaxies (identified by solid red points in the plots)
may be steeper than those of ``central'' galaxies, consistent with the
picture of tidal truncation (Figure~\ref{fig:histonbobjff1}).  The
average $f_{\rm SIE}$ is larger for ``satellite'' galaxies than for
``central'' galaxies (approximately at 95\%CL). However, a K-S test
shows that the probability that the two distributions are drawn from
the same parent distribution is non-negligible (7.5\%) and therefore
we caution that this difference may not be significant. In this sense,
and in terms of overall significance, our finding agrees with that
reported by \citet{Aug08} who studied the initial SLACS sample of 15
lenses, finding tentative evidence for steeper mass density profiles
in overdense regions.  As we discuss in the next section, an
alternative or possibly complementary explanation is that $f_{\rm
SIE}$ is somewhat higher for the central galaxies as a result of an
additional contribution to the lensing mass by the dark-matter halo of
the overdensity.

\subsection{Enclosed mass, external convergence and the average mass-density structure of early-type galaxies}
\label{ssec:ML}

Finally, we study the radial dependence of mass enclosed within the
Einstein radius as a function of local and global environment.  As
discussed in Paper~VII, for an isolated galaxy described by a singular
isothermal sphere total mass density profile, we expect the following
relation:

\begin{equation}
\log \frac{M(<R)}{M_{\rm dim}} = \log \frac{R}{R_{\rm e}} + \log 2 c_{\rm e2},
\label{eq:miso}
\end{equation}

\noindent
where M$_{\rm dim}=\sigma_{\rm e,2}^2 R_{\rm e}/2 G $ is the
``dimensional'' mass, $\sigma_{\rm e,2}$ is the stellar velocity
dispersion corrected to an aperture of radius equal to half the
effective radius $R_{\rm e}$, $G$ is the gravitational constant, and
$\cet \equiv \frac{2GM_{\rm e2}}{R_{\rm e}\sigma_{\rm e,2}^2}$ --
where $M_{\rm e2}$ is the projected mass enclosed by a circle of
radius equal to one half the effective radius -- is a dimensionless
structure parameter that depends on the profile and anisotropy of the
luminous tracer \citep{NTB08}. As shown in Figure~\ref{fig:M1},
Equation~\ref{eq:miso} is found to be a good description of the SLACS
sample, with an intercept $\log 2\cet=0.83\pm0.01$.  For typical
luminosity profiles, such as \citet{deV48} with close to isotropic
pressure tensor, $\log \cet \approx 0.57$ \citep{NTB08}, consistent
with the best fit value of the intercept of
Equation~\ref{eq:miso}. The dimensionless structure parameter can vary
by as much as $\sim0.05$ dex within the realm of physically plausible
galaxy models, reproducing the intrinsic scatter derived by
\citet{Bol++08b} on the basis of the mass plane analysis.

This description of the SLACS sample is approximately equivalent of
that in the terms of the mass plane (MP; i.e. the correlation between
stellar velocity dispersion, effective radius and surface mass
density) presented by \citet{Bol++07,Bol++08b} and \citet{NTB08},
since Eq~\ref{eq:miso} evaluated at $R=R_{\rm e}/2$ returns an
equation very close to that describing the mass plane.

We can use the scaling relation in Eq~\ref{eq:miso} to measure the
effects of the environment in the form of an additional surface mass
density\footnote{To avoid duplication, we only describe the trends of
Equation~\ref{eq:miso} with environment, the trends of the MP being
effectively indistinguishable and less straightforward to interpret.},
i.e. external convergence in the language of gravitational lensing.
If the lens galaxy can be described as an isothermal sphere plus a
uniform surface mass density representing the first order expansion of
the mass distribution along the line of sight from material not
associated with the lens, the surface mass density profile of the lens
becomes:

\begin{equation}
\kappa=\frac{1}{2}\frac{R_{\rm Ein,0}}{R_{\rm e}}+\kappa_{\rm ext},
\label{eq:kappa}
\end{equation}

\noindent
where the surface mass density $\kappa$ is expressed as usual
\cite[e.g.,][]{Sch06} in units of the critical density $\Sigma_{\rm
crit}=\frac{c^2 D_s}{4\pi G D_d D_{ds}}$ ($c$ is the speed of light,
$D_s$, $D_d$, $D_{ds}$ are the angular diameter distances from the
observer to the source, from the observer to the lens, and between the
lens and the source, respectively). R$_{\rm Ein,0}$ is the Einstein
radius that the galaxy would have if there was no external surface
mass density contribution $\kappa_{\rm ext}$.

If we were to neglect the presence of the external surface mass
density and model the galaxy as an isolated isothermal sphere, we
would infer the following Einstein radius and Einstein mass:

\begin{equation}
R_{\rm Ein}=\frac{R_{\rm Ein,0}}{1-\kappa_{\rm ext}};
M_{\rm Ein}=\frac{M_{\rm Ein,0}}{(1-\kappa_{\rm ext})^2},
\label{eq:ms}
\end{equation}

\noindent
expressed in units of the corresponding quantities for the isolated
lens.  Using lensing alone we would not be able to detect the presence
of the external surface mass density, due to the mass-sheet degeneracy
\citep[][]{FGS85,Koc06}. However, the external surface mass density
modifies Equation~\ref{eq:miso}:

\begin{equation}
\log \frac{M_{\rm Ein}}{M_{\rm dim}} = \log \frac{R_{\rm Ein}}{R_{\rm e}} + \log 2 c_{\rm e2}-\log(1-\kappa_{\rm ext}),
\label{eq:misokappa}
\end{equation}

\noindent obtained by combining Equations~\ref{eq:ms}
and~\ref{eq:miso}.  Therefore we can in principle measure the external
mass distribution using this scaling law\footnote{This procedure is
similar to that adopted by \citet{Koo++03} and Suyu et al.\ (2008) to
break the mass-sheet degeneracy in measuring the Hubble constant by
relying on stellar velocity dispersion measurements.}. Vertical
offsets from the average relation in Figure~\ref{fig:M1}

\begin{equation}
\Delta \log \frac{M_{\rm Ein}}{M_{\rm dim}}= \log \frac{M_{\rm Ein}}{M_{\rm dim}}-\log \frac{R_{\rm Ein}}{R_{\rm e}}-0.83,
\label{eq:deltam}
\end{equation}

\noindent can be interpreted as due to the environment if $\cet$ is
constant. Note however that there is a degeneracy between the
dimensionless structure parameter $\cet$ and $\kappa_{\rm ext}$, which
should be kept in mind while interpreting any trend.

To quantify any trends we show the offset from the average relation as
a function of local and global overdensity in Figure~\ref{fig:M3}. No
significant correlation is found, suggesting that any environmental
effect is smaller than our measurement errors. The slope of the
relations $\Delta \log M_{\rm Ein} / M_{\rm dim} =
\eta_{1,\Sigma_{10}} \log \Sigma_{10}/\langle
\Sigma_{10}\rangle+\eta_{0,\Sigma_{10}}$ and $\Delta \log M_{\rm Ein}
/ M_{\rm dim} = \eta_{1,D_1} \log D_{1}/\langle
D_{1}\rangle+\eta_{0,D_1}$ are consistent with zero:
$\eta_{1,\Sigma_{10}}=-0.01\pm0.03$ and
$\eta_{1,D_{1}}=-0.07\pm0.05$. Thus, even in the highest density
environments ($\Sigma_{10}/\langle \Sigma_{10}\rangle\sim100$;
$D_1/\langle D_1\rangle\sim10$), we do not expect the external
contribution to the total mass to be more than 0.03 dex and 0.06 dex
respectively, at the 95\% CL, corresponding to external mass surface
densities of less than $\kappa_{\rm ext}\sim 0.07$ and $\kappa_{\rm
ext}\sim 0.13$, assuming that the dimensionless structure parameter is
independent of environment.

Considering special subclasses of objects, the satellite galaxies are
below the average ($\langle\Delta M_{\rm Ein}/M_{\rm dim}\rangle_{\rm
sat}=-0.07\pm0.03$ dex; the difference with respect to central
galaxies being $-0.09\pm0.03$), while the galaxies in known clusters
are consistent with the average ($\langle \Delta M_{\rm Ein}/M_{\rm
dim}=0.02\pm0.02 \rangle_{\rm clu}$).  These offsets could be
interpreted as a combination of two effects: i) negative external mass
contributions for the satellites (negative $\kappa_{\rm ext}$), i.e.\
satellite galaxies living in underdense regions; ii) different $\cet$
for satellite and central galaxies. Consistent with the analysis of
the results presented in \S~\ref{ssec:ff}, it seems that the most
plausible interpretation of the offset between the satellite and
central galaxies is the latter. This could arise for example from
systematic changes of their internal structure due to tidal stripping
of their dark-matter halos. However, a combination of the two effects
cannot be ruled out without more information, e.g. from spatially
resolved stellar kinematics \citep{Czo++08}.

The intrinsic scatter around the best fit relation (0.059 dex along
the vertical axis) is also of interest, because it allows us to set
limits to the combined degree of inhomogeneity in the internal
structure of the lens galaxies and on the external mass surface
density. \citet{NTB08} used the small internal scatter of the relation
to constrain $\log \cet$ and thus the internal structure of the
lenses, neglecting the effects of external mass along the line of
sight. Considering that part of the scatter may be due to fluctuations
in external mass along the line of sight it is likely that this is an
upper limit to degree of inhomogeneity of the internal structure of
early-type galaxies. Conversely, if we were to postulate that lens
galaxies are absolutely homogeneous in their internal structure, the
observed internal structure would imply an r.m.s. scatter of 0.05 dex
in $\log (1-\kappa_{\rm ext})$, i.e. $\sim0.1$ in $\kappa_{\rm
ext}$. We can therefore take this number as an upper limit to the
r.m.s contribution of the external potential, in line with the
expectations of theoretical calculations for image separations of
order one arcsecond
\citep{OKD05}.

\section{Discussion}
\label{sec:disc}

Our study addresses two issues which have gathered significant
attention in the literature in the past few years: i) the role of the
environment in the modeling and interpretation of gravitational
lensing results; ii) the role of environment in shaping the structure
and dynamics of galaxies. 

As far as the first issue is concerned, we find some evidence that
lens galaxies tend to live preferentially in overdense regions,
although the scatter is large. As pointed out by a number of authors,
this is expected since gravitational lenses are typically massive
galaxies and are therefore clustered.  In fact, one of the main
results of this paper is that the distribution of environments for
lens galaxies and ``twins'' selected to have the same velocity
dispersions and redshifts are indistinguishable within the
errors. This conclusion extends our previous finding that {\it SLACS
lenses are normal galaxies} that just happen to be well aligned with a
background source.

Defining membership in groups/clusters either by the association of
lens galaxies with group/clusters known to NED or by (somewhat
arbitrary) cuts in overdensity ($D_1/\langle D_1\rangle>1$ and
$\Sigma_{10}/\langle\Sigma_{10}\rangle>2$) we find that approximately
20\% of the SLACS lenses belong to a known group/cluster
($12/70=17\pm$5\% and $13/66=20\pm6$\%, respectively, according to the
two definitions). This is in line with the theoretical estimates by
\citet{KCZ00} and somewhat lower than suggested by spectroscopic and
photometric studies of other samples of lenses, typically at higher
redshift \citep{F+L02,Mom++06,Aug++07,Fas++07,Wil++08}, although a
direct comparison of the fraction is difficult because of the
different definitions of groups/clusters adopted by various studies.

In addition, we find that the contribution of the environment to the
potential of the SLACS lenses is small, typically undetected, reaching
10-20 \% in surface mass density (corresponding to external shear and
convergence of order 0.05-0.10) only for a few extreme cases. This is
consistent with our previous studies that showed that SLACS lenses can
be successfully modeled by singular isothermal spheres without
external shear, and at variance with other sample of lenses where
substantial amounts of external shear are required for successful
models \citep[e.g.][]{H+B94,KKS97,Mou++07}. Our result is also in
disagreement with the study by \citet{G+S07b} who suggested
significant external convergence for the SLACS sample based on simple
one-component models (i.e. light traces mass), which are also ruled
out by a number of other arguments as discussed in previous SLACS
papers.

The reduced role of the environment in the SLACS lenses with respect
to other samples of lenses is likely to be due to the lower redshift
of lens and source and consequently to the smaller Einstein radii both
in terms of angular scales ($\langle {\rm R}_{\rm Ein} \rangle =
1\farcs2$) and in terms of physical scales of the lens galaxies,
typically half the effective radius as opposed, e.g., to a few
effective radii for the lenses studied by the LSD project
\citep{T+K04} and the typical CASTLES lenses \citep{Rus++03}. In
addition, the SDSS fiber selection imposes effectively an upper cutoff
to the image separation, corresponding to $\lesssim 2''$ \citep[see
discussion in][]{Dob++08}, in order for a sizeable fraction of the
lensed flux to be captured by the spectrograph.  The two effects work
in the same direction, increasing the relative importance of the
stellar mass for the lens model with respect with other sample of
lenses and ruling out the largest image separation lenses where the
effects of group and cluster environments are expected to be largest
\citep{OKD05}.  The low level of environmental contribution to the
lens potential is a major benefit of the SLACS sample, reducing the
risk of biased inferences on the structure of the lenses due to poor
modeling of the environment \citep{K+Z04}. For example, if time
variable phenomena are identified in the SLACS lensed sources
(e.g. active nuclei or supernovae) they would make excellent systems
to determine the Hubble constant from gravitational time delays.

The environment and the distortion caused by external shear has often
been considered one of the potential solutions to the so-called ``quad
problem'' \citep{R+T01,H+S03}, i.e. the abundance of the quadruply
imaged lenses amongst radio-selected lenses (9/22 in the CLASS survey,
plus one system with six images; \cite{Bro++03}). The ``quad-problem''
is not observed for optically selected multiply imaged quasars
\citep{Ogu07a}, suggesting that the faint end slope of the 
luminosity function of the sources is different for radio and
optically selected sources. The classification in doubles and quads is
not so clear cut for SLACS, because of the extended nature and
sometimes multiple components of the sources. However, if we base our
classification on the criterion that the peak of the brightest source
being inside or ``on'' the inner caustic, only 9/70 ($13\pm5$\%) SLACS
lenses are classified as quads\footnote{The lenses were classified
independently by T.T. and A.S.B.. The nine ``consensus'' quads are:
J0405-0455, J0737+3216, J0912+0029, J0946+1006, J0956+5100,
J1100+5329, J1106+5228, J1402+6321, J1420+6019, J2300+0022. Three
additional lenses are classified as quads by only one of two
classifiers, J0956+5100, J1103+5322, J2341+0000, possibly increasing
the fraction to 12/70, i.e. $17\pm5$\%}, significantly lower than the
radio selected samples. Qualitatively, this is in line with the
expectations of models that do not require large amounts of external
shear \citep[e.g.,][]{R+T01,Ogu07a}. However, a full modeling of the
SLACS source distribution and selection function is required for a
quantitative comparison. This is left for future work, when we will
analyze the properties of the source galaxies in more detail.

As far as the second issue is concerned, we do not find significant
correlations between the internal properties of the lens galaxies and
the environment. The only difference, detected at 2-3 $\sigma$
significance, is between central and satellite galaxies, in the sense
that the latter have somewhat steeper mass density profiles than the
former do. We argue that this may be interpreted as due to tidal
truncation of the outer parts of the galaxies, as suggested by
numerical \citep{DKF07} and observational studies
\citep{Imp++98,T+K02b,Aug08}. This finding extends at smaller scales previous 
evidence for tidal truncation at larger radii, based on the weak
galaxy-galaxy lensing signal \citep{Nat++02,Gav++04,Nat++07}.
However, current data cannot exclude the possibility that this trend
is in part due to an extra amount of convergence from the larger-scale
structure around central galaxies. In the future, we hope to break
this degeneracy with high quality spatially resolved kinematics. In
any case, the small intrinsic scatter in the slope of the mass density
profiles and around Equation~\ref{eq:miso} (or equivalently the MP)
shows that SLACS lenses are both highly homogeneous in their internal
structure and suffer from little contamination by large scale
structures along the line of sight.

A limitation of this study is the reliance on photometric catalog of
galaxies as tracers of the environment.  X-ray data and spectroscopic
redshifts for a substantial fraction of the objects in each field are
needed to make further progress, by providing accurate position and
masses for the groups and clusters in the vicinity of the lens and
therefore an independent estimate of the external convergence and
shear. X-ray data in particular would be needed to detect fossil
groups.

\section{Summary and conclusions}

\label{sec:sum}

We have used the SDSS database to measure the environment around
seventy strong gravitational lens galaxies selected from the SLACS
Survey.  We adopt two standard estimators: the surface density of
galaxies within the tenth nearest neighbor ($\Sigma_{10}$) and the
density of galaxies within a cone of radius one $h^{-1}$Mpc
($D_1$). Both are normalized in terms of the corresponding quantities
for random fields, and are referred to as ``local'' and ``global''
overdensities. For comparison purposes, we also selected from the SDSS
database a sample of 100 ``twins'' for each lens galaxy, i.e.\
galaxies with virtually the same redshift and velocity dispersion. The
new observables are combined with measurements of internal properties
of the lens galaxies from SLACS papers to investigate the relationship
between early-type galaxy structure and environment. The main results
of this study can be summarized as follows:

\begin{enumerate}
\item SLACS lens galaxies appear to live generally in somewhat 
overdense environments. Twelve of the seventy lenses ($17\pm5$\%) are
associated with known clusters/groups at the same redshifts. This is
consistent with the notion that lens galaxies are massive and
therefore clustered.

\item The distribution of overdensities for SLACS lens galaxies is
consistent within the errors with the corresponding distribution for
``twin'' non-lens galaxies. This is consistent with lens galaxies
being an unbiased population of massive early-type galaxies with
respect to their environment.

\item The misalignment of mass and light is found to correlate 
with the local overdensity of galaxies. The misalignment is negligible
for most lens galaxies except for those in the most overdense
regions. Randomly oriented external shear of order of 0.05-0.06 is
required to reproduce the observed misalignment in the most overdense
environments.

\item The small departures from the average relation between Einstein mass,
scaled by dimensional mass, and Einstein radius, scaled by the
effective radius, support the previous conclusions. The contribution
of the environment to the local potential of the main lens is
estimated to be below the current detection threshold for most lenses
except for those residing in the densest environments where it can
reach at most 10-20\% in terms of surface mass density at the Einstein
radius (0.05-0.1 external convergence).

\item No significant correlation is found between local and global
overdensity and measures of internal structure, such as the slope of
the total mass density profile--- quantified in terms of $f_{\rm
SIE}$--- and the difference between the observed Einstein mass and
that predicted based on dimensional mass, effective radius, and
Einstein radius. Thus -- within the current level of precision -- the
internal structure of early-type galaxies does not appear to be biased
by projection effects. 

\item The properties of ``satellite'' galaxies (i.e., those with a
nearby companion with $i'<i'_{\rm lens}-1$) are found to be different
than those of ``central'' galaxies at the 95-99\% CL. The average
slope of the total mass density profile is $f_{\rm SIE}=1.12\pm0.05$
for the satellites and $1.01\pm0.01$ for the central
galaxies. Similarly, the ratio between Einstein Mass and M$_{\rm dim}$
--- as a function of the Einstein radius in units of the effective
radius ratio --- is $0.09\pm0.03$ dex lower for the satellites with
respect to the central galaxies. This suggests that the outer parts of
satellite galaxies are perturbed by the environment down to the kpc
scales probed by strong lensing, consistent with tidal stripping.
\end{enumerate}

\acknowledgments

We thank Steve Allen, Matt Auger, Maru\v{s}a Brada\v{c}, and Chris
Fassnacht for many insightful conversations.  Support for programs
\#10174, \#10587, \#10886, \#10494, \#10798 was provided by NASA through 
a grant from the Space Telescope Science Institute, which is operated
by the Association of Universities for Research in Astronomy, Inc.,
under NASA contract NAS 5-26555.  T.T.  acknowledges support from the
NSF through CAREER award NSF-0642621, by the Sloan Foundation through
a Sloan Research Fellowship, and by the Packard Foundation through a
Packard Fellowship.  L.V.E.K. is supported (in part) through an
NWO-VIDI program subsidy (project number 639.042.505). The work of
L.A.M.  was carried out at the Jet Propulsion Laboratory, California
Institute of Technology, under a contract with NASA.  This research
has made use of the NASA/IPAC Extragalactic Database (NED) which is
operated by the Jet Propulsion Laboratory, California Institute of
Technology, under contract with the National Aeronautics and Space
Administration. This project would not have been feasible without the
extensive and accurate database provided by the Digital Sloan Sky
Survey (SDSS).  Funding for the creation and distribution of the SDSS
Archive has been provided by the Alfred P. Sloan Foundation, the
Participating Institutions, the National Aeronautics and Space
Administration, the National Science Foundation, the U.S. Department
of Energy, the Japanese Monbukagakusho, and the Max Planck
Society. The SDSS Web site is http://www.sdss.org/.  The SDSS is
managed by the Astrophysical Research Consortium (ARC) for the
Participating Institutions. The Participating Institutions are The
University of Chicago, Fermilab, the Institute for Advanced Study, the
Japan Participation Group, The Johns Hopkins University, the Korean
Scientist Group, Los Alamos National Laboratory, the
Max-Planck-Institute for Astronomy (MPIA), the Max-Planck-Institute
for Astrophysics (MPA), New Mexico State University, University of
Pittsburgh, University of Portsmouth, Princeton University, the United
States Naval Observatory, and the University of Washington.

%\bibliographystyle{apj}
%\bibliography{litter/references}

\clearpage
%\LongTables

\begin{deluxetable}{lrrrrrrrrlrr}
%\rotate
\setlength{\tabcolsep}{3pt}
\tabletypesize{\scriptsize}
%\tablewidth{0}
\tablecaption{Summary of relevant measurements}
\tablehead{
\colhead{Lens}        &
\colhead{$z_l$}        &
\colhead{$\sigma_*$}        &
\colhead{$\sigma_{\rm SIE}$}        &
\colhead{$\Sigma_{10}$}     &
\colhead{$\langle\Sigma_{10}\rangle_t$}           &
\colhead{$D_1$}     &
\colhead{$\langle D_1\rangle$}    &
\colhead{$\langle D_1\rangle_t$}           &
\colhead{Cluster Name} &
\colhead{$z_{cl}$} &
\colhead{Dist} \\
\colhead{(1)} &
\colhead{(2)} &
\colhead{(3)} &
\colhead{(4)} &
\colhead{(5)} &
\colhead{(6)} &
\colhead{(7)} &
\colhead{(8)} &
\colhead{(9)} &
\colhead{(10)} &
\colhead{(11)} &
\colhead{(12)} }
\tablecolumns{12}
\startdata
J0008$-$0004 & 0.440 &     $-$     & 271 & 3.8$\pm$1.2 & 10.1$\pm$2.1 & 3.1$\pm$0.7 & 4.2$\pm$0.1 & 4.4$\pm$0.3 & SDSSCEJ002.090289-00.168613 & 0.254 & 7.60 \\
J0029$-$0055 & 0.227 & 245$\pm$19  & 217 & 3.3$\pm$1.0 & 6.5$\pm$0.7 & 2.7$\pm$0.6 & 3.8$\pm$0.1 & 3.8$\pm$0.2  & None & $-$ & $-$  	\\
J0037$-$0942 & 0.195 & 299$\pm$15  & 285 & 1.8$\pm$0.6 & 8.4$\pm$1.2 & 1.7$\pm$0.5 & 3.0$\pm$0.2 & 3.8$\pm$0.2  & None & $-$ & $-$  	\\
J0044$+$0113 & 0.120 & 283$\pm$14  & 269 &     $-$     &     $-$     &     $-$     &     $-$     &     $-$      & ZwCl0041.9+0052 & $-$ & 7.90 \\
J0109$+$1500 & 0.294 & 274$\pm$21  & 243 & 4.5$\pm$1.4 & 6.4$\pm$0.7 & 3.1$\pm$0.7 & 4.0$\pm$0.1 & 3.5$\pm$0.1  & None & $-$ & $-$  	\\
J0157$-$0056 & 0.513 &     $-$     & 269 & 12.1$\pm$3.8 & 6.4$\pm$0.7 & 8.3$\pm$1.1 & 3.5$\pm$0.1 & 3.4$\pm$0.1 & SDSSCEJ029.619265-00.943160 & 0.424 & 7.40 \\
J0216$-$0813 & 0.332 & 354$\pm$24  & 347 & 9.5$\pm$3.0 & 7.6$\pm$1.0 & 3.1$\pm$0.7 & 1.6$\pm$0.1 & 4.0$\pm$0.2  & None & $-$ & $-$  	\\
J0252$+$0039 & 0.280 & 179$\pm$13  & 235 & 2.3$\pm$0.7 & 7.6$\pm$1.1 & 1.7$\pm$0.5 & 5.0$\pm$0.1 & 4.7$\pm$0.3  & None & $-$ & $-$  \\
J0330$-$0020 & 0.351 & 232$\pm$23  & 252 & 11.6$\pm$3.7 & 6.4$\pm$0.5 & 4.8$\pm$0.9 & 4.3$\pm$0.1 & 4.1$\pm$0.2 & None & $-$ & $-$  \\
J0405$-$0455 & 0.075 & 175$\pm$9   & 177 & 3.8$\pm$1.2 & 6.4$\pm$0.7 & 3.7$\pm$0.8 & 7.3$\pm$0.4 & 3.5$\pm$0.1  & None & $-$ & $-$  	\\
J0728$+$3835 & 0.206 & 231$\pm$12  & 256 & 5.1$\pm$1.6 & 7.9$\pm$0.9 & 4.1$\pm$0.8 & 3.8$\pm$0.2 & 4.3$\pm$0.2  & None & $-$ & $-$  	\\
J0737$+$3216 & 0.322 & 358$\pm$18  & 292 & 3.4$\pm$1.1 & 7.7$\pm$1.1 & 2.7$\pm$0.6 & 2.2$\pm$0.1 & 4.0$\pm$0.2  & None & $-$ & $-$  	\\
J0808$+$4706 & 0.220 &     $-$     & $-$ & 1.2$\pm$0.4 & 8.1$\pm$1.0 & 1.1$\pm$0.4 & 2.3$\pm$0.1 & 3.8$\pm$0.2  & None & $-$ & $-$  	\\
J0822$+$2652 & 0.241 & 279$\pm$16  & 271 & 2.1$\pm$0.7 & 9.4$\pm$1.1 & 2.0$\pm$0.6 & 3.4$\pm$0.1 & 4.2$\pm$0.2  & None & $-$ & $-$  	\\
J0841$+$3824 & 0.116 & 235$\pm$11  & 248 & 3.2$\pm$1.0 & 10.2$\pm$1.5 & 2.7$\pm$0.6 & 2.8$\pm$0.2 & 4.7$\pm$0.3 & MaxBCGJ130.27341+38.53306 & 0.240 & 9.00 \\
J0903$+$4116 & 0.430 &     $-$     & 293 & 24.9$\pm$7.9 & 6.8$\pm$0.6 & 5.8$\pm$0.9 & 2.7$\pm$0.1 & 3.6$\pm$0.1 & None & $-$ & $-$  \\
J0912$+$0029 & 0.164 & 341$\pm$17  & 346 & 5.5$\pm$1.7 & 12.1$\pm$1.8 & 3.3$\pm$0.7 & 2.5$\pm$0.1 & 4.3$\pm$0.2 & None & $-$ & $-$  \\
J0935$-$0003 & 0.347 & 413$\pm$36  & 361 & 8.3$\pm$2.6 & 9.4$\pm$1.1 & 3.7$\pm$0.8 & 1.4$\pm$0.1 & 4.5$\pm$0.2  & None & $-$ & $-$   \\
J0936$+$0913 & 0.190 & 260$\pm$13  & 243 & 4.8$\pm$1.5 & 9.3$\pm$1.3 & 4.8$\pm$0.9 & 3.5$\pm$0.2 & 4.4$\pm$0.3  & MaxBCGJ143.87624+09.27139 & 0.127 & 8.00 \\
J0946$+$1006 & 0.222 & 281$\pm$22  & 283 & 3.4$\pm$1.1 & 7.4$\pm$0.9 & 3.0$\pm$0.7 & 4.1$\pm$0.2 & 3.9$\pm$0.2  & MaxBCGJ146.87912+10.07800 & 0.151 & 8.70 \\
J0955$+$0101 & 0.111 & 211$\pm$14  & 224 & 7.7$\pm$2.4 & 10.5$\pm$2.7 & 4.8$\pm$0.9 & 5.4$\pm$0.3 & 4.6$\pm$0.5 & SDSSCEJ148.948990+01.050020 & 0.311 & 7.10 \\
J0956$+$5100 & 0.241 & 358$\pm$18  & 318 & 5.0$\pm$1.6 & 7.9$\pm$0.9 & 4.5$\pm$0.8 & 2.4$\pm$0.1 & 3.6$\pm$0.2  & MaxBCGJ149.12403+51.00178 & 0.235 & 0.00  \\
J0959$+$4416 & 0.237 & 262$\pm$20  & 254 & 2.8$\pm$0.9 & 7.4$\pm$0.9 & 2.5$\pm$0.6 & 3.4$\pm$0.1 & 3.9$\pm$0.2  & None & $-$ & $-$  	\\
J0959$+$0410 & 0.126 & 215$\pm$14  & 216 & 17.5$\pm$5.5 & 6.8$\pm$0.8 & 7.2$\pm$1.1 & 5.3$\pm$0.3 & 4.7$\pm$0.3 & NSCJ095952+040356 & 0.153 & 6.70 \\
J1016$+$3859 & 0.168 & 269$\pm$14  & 253 & 8.4$\pm$2.6 & 7.4$\pm$0.9 & 10.1$\pm$1.3 & 4.4$\pm$0.2 & 3.9$\pm$0.2 & NSCJ101706+390221 & 0.206 & 8.60 \\ 	
           &       &             &     &             &             &             &             &              & ABELL0963 & 0.206 & 9.30 \\
           &       &             &     &             &             &             &             &              & NSCSJ101645+385041 & 0.230 & 9.40 \\ 
J1020$+$1122 & 0.282 & 306$\pm$20  & 304 & 3.2$\pm$1.0 & 7.1$\pm$0.8 & 2.5$\pm$0.6 & 3.2$\pm$0.1 & 3.9$\pm$0.2  & None & $-$ & $-$  	\\
J1023$+$4230 & 0.191 & 261$\pm$16  & 267 & 6.1$\pm$1.9 & 8.5$\pm$1.1 & 4.2$\pm$0.8 & 4.0$\pm$0.2 & 4.1$\pm$0.2  & MaxBCGJ155.91525+42.48492 & 0.184 & 1.70 \\
J1029$+$0420 & 0.105 & 228$\pm$12  & 209 & 2.6$\pm$0.8 & 8.5$\pm$1.1 & 3.0$\pm$0.7 & 4.3$\pm$0.3 & 4.1$\pm$0.2  & None & $-$ & $-$  	\\
J1032$+$5322 & 0.133 & 330$\pm$17  & 250 & 11.0$\pm$3.5 & 10.6$\pm$1.3 & 9.7$\pm$1.2 & 5.7$\pm$0.3 & 4.4$\pm$0.3 & MaxBCGJ158.03906+53.32018 & 0.138 & 5.20 \\
J1100$+$5329 & 0.317 &     $-$     & 303 & 3.5$\pm$1.1 & 6.3$\pm$0.6 & 3.1$\pm$0.7 & 1.9$\pm$0.1 & 4.1$\pm$0.2  & None & $-$ & $-$   \\
J1103$+$5322 & 0.158 & 211$\pm$13  & 217 & 3.9$\pm$1.2 & 7.5$\pm$0.7 & 3.0$\pm$0.7 & 4.0$\pm$0.2 & 4.7$\pm$0.2  & None & $-$ & $-$  	\\
J1106$+$5228 & 0.095 & 283$\pm$14  & 239 & 2.1$\pm$0.7 & 10.8$\pm$2.0 & 1.7$\pm$0.5 & 3.5$\pm$0.2 & 4.9$\pm$0.5 & NSCSJ110634+522247 & $-$ & 6.10 \\
J1112$+$0826 & 0.273 & 348$\pm$22  & 314 & 4.1$\pm$1.3 & 8.3$\pm$1.4 & 2.2$\pm$0.6 & 3.1$\pm$0.1 & 3.8$\pm$0.2  & None & $-$ & $-$  	\\
J1134$+$6027 & 0.153 & 257$\pm$13  & 242 & 6.6$\pm$2.1 & 11.9$\pm$1.8 & 4.4$\pm$0.8 & 4.2$\pm$0.2 & 5.1$\pm$0.3 & None & $-$ & $-$  \\
J1142$+$1001 & 0.222 & 238$\pm$24  & 254 & 2.6$\pm$0.8 & 8.5$\pm$1.0 & 1.4$\pm$0.5 & 3.6$\pm$0.2 & 4.5$\pm$0.2  & None & $-$ & $-$  	\\
J1143$-$0144 & 0.106 & 279$\pm$13  & 285 & 237$\pm$75  & 12.5$\pm$1.6 & 11.2$\pm$1.3 & 3.0$\pm$0.2 & 4.7$\pm$0.5 & SDSS-C41035 & 0.106 & 0.90 \\
           &       &             &     &             &             &             &             &              & ABELL1364 & 0.106 & 2.70    \\
           &       &             &     &             &             &             &             &              & MaxBCGJ176.02643-01.79024 & 0.257 & 9.60 \\ 
J1153$+$4612 & 0.180 & 248$\pm$16  & 220 & 3.4$\pm$1.1 & 8.3$\pm$0.9 & 3.4$\pm$0.7 & 5.2$\pm$0.2 & 4.7$\pm$0.3  & None & $-$ & $-$  	\\
J1204$+$0358 & 0.164 & 290$\pm$18  & 254 & 37$\pm$12   & 8.8$\pm$1.4 & 15.9$\pm$1.6 & 4.7$\pm$0.2 & 4.3$\pm$0.3 & MaxBCGJ181.14640+03.95642 & 0.165 & 2.30  \\
           &       &             &     &             &             &             &             &              & NSCJ120432+035012 & 0.116 & 8.40 \\
           &       &             &     &             &             &             &             &              & MaxBCGJ181.20101+03.98553 & 0.265 & 1.50\\
           &       &             &     &             &             &             &             &              & ABELL1463 & $-$ & 1.90 \\
J1205$+$4910 & 0.215 & 299$\pm$15  & 285 & 2.3$\pm$0.7 & 7.2$\pm$1.1 & 2.2$\pm$0.6 & 2.8$\pm$0.1 & 3.6$\pm$0.2  & None & $-$ & $-$  	\\
J1213$+$6708 & 0.123 & 308$\pm$16  & 251 & 1.5$\pm$0.5 & 8.6$\pm$0.9 & 1.4$\pm$0.5 & 3.1$\pm$0.2 & 3.6$\pm$0.2  & None & $-$ & $-$  	\\
J1218$+$0830 & 0.135 & 231$\pm$12  & 254 & 1.4$\pm$0.4 & 9.2$\pm$1.2 & 1.2$\pm$0.4 & 3.2$\pm$0.2 & 4.5$\pm$0.2  & None & $-$ & $-$  	\\
J1250$+$0523 & 0.232 & 272$\pm$15  & 244 & 3.1$\pm$1.0 & 7.0$\pm$0.6 & 2.8$\pm$0.7 & 2.8$\pm$0.1 & 3.9$\pm$0.2  & None & $-$ & $-$  	\\
J1250$-$0135 & 0.087 & 260$\pm$13  & $-$ & 5.8$\pm$1.8 & 12.4$\pm$2.2 & 6.7$\pm$1.0 & 3.1$\pm$0.2 & 4.5$\pm$0.4 & None & $-$ & $-$  \\
J1251$-$0208 & 0.224 &     $-$     & 209 & 7.8$\pm$2.5 & 9.4$\pm$1.2 & 3.1$\pm$0.7 & 4.1$\pm$0.2 & 4.3$\pm$0.2  & NSCJ125151-021711 & 0.169 & 10.00 \\
J1259$+$6134 & 0.233 & 273$\pm$17  & $-$ & 2.3$\pm$0.7 & 9.4$\pm$1.2 & 1.9$\pm$0.5 & 3.2$\pm$0.1 & 4.3$\pm$0.2  & None & $-$ & $-$  	\\
J1402$+$6321 & 0.205 & 283$\pm$18  & 294 & 1.8$\pm$0.6 & 9.8$\pm$1.1 & 1.7$\pm$0.5 & 2.7$\pm$0.2 & 4.3$\pm$0.3  & NSCSJ140131+632201 & 0.350 & 6.40 \\
J1403$+$0006 & 0.189 & 232$\pm$18  & 225 & 31.1$\pm$9.8 & 8.6$\pm$1.0 & 10.6$\pm$1.3 & 4.6$\pm$0.2 & 4.6$\pm$0.3 & SDSSCEJ210.802505+00.093432 & 0.183 & 4.40 \\ 
           &       &             &     &             &             &             &             &               & MaxBCGJ210.85765+00.13363 & 0.167 & 1.60 \\
J1416$+$5136 & 0.299 & 261$\pm$27  & 287 & 3.9$\pm$1.2 & 7.2$\pm$0.9 & 3.4$\pm$0.7 & 3.4$\pm$0.1 & 3.8$\pm$0.2   & None & $-$ & $-$  	\\
J1420$+$6019 & 0.063 & 220$\pm$11  & 204 & 1.8$\pm$0.6 & 10.4$\pm$1.4 & 1.7$\pm$0.5 & 5.1$\pm$0.4 & 5.4$\pm$0.4  & None & $-$ & $-$  \\
J1430$+$4105 & 0.285 & 343$\pm$34  & 337 & 3.3$\pm$1.1 & 9.6$\pm$1.2 & 2.2$\pm$0.6 & 1.8$\pm$0.1 & 4.0$\pm$0.3   & MaxBCGJ217.49493+41.10435 & 0.270 & 1.00 \\ 
J1432$+$6317 & 0.123 & 205$\pm$10  & 236 &     $-$     &     $-$     &     $-$     &     $-$     &     $-$       & None & $-$ & $-$  	\\
J1436$-$0000 & 0.285 & 240$\pm$18  & 256 & 5.6$\pm$1.8 & 6.3$\pm$0.6 & 4.4$\pm$0.8 & 2.9$\pm$0.1 & 3.4$\pm$0.2   & SDSSCEJ219.075027+00.094154 & 0.288 & 6.60 \\ 	
           &       &             &     &             &             &             &             &               & SDSSCEJ218.997833-00.116972 & 0.139 & 9.60 \\
J1443$+$0304 & 0.134 & 231$\pm$12  & 207 & 4.9$\pm$1.5 & 6.7$\pm$0.7 & 4.8$\pm$0.9 & 5.9$\pm$0.3 & 4.4$\pm$0.3   & None & $-$ & $-$  	\\
J1451$-$0239 & 0.125 & 238$\pm$15  & 222 & 3.8$\pm$1.2 & 7.8$\pm$0.8 & 3.9$\pm$0.8 & 3.6$\pm$0.2 & 4.4$\pm$0.3   & ZwCl1449.1-0227 & $-$ & 3.40 \\
J1525$+$3327 & 0.358 & 279$\pm$28  & 318 & 1.2$\pm$0.4 & 8.3$\pm$1.0 & 1.1$\pm$0.4 & 1.9$\pm$0.1 & 4.3$\pm$0.2   & NSCJ152503+332621 & 0.219 & 1.60 \\
J1531$-$0105 & 0.160 & 297$\pm$15  & 281 & 4.8$\pm$1.5 & 8.9$\pm$0.8 & 4.4$\pm$0.8 & 3.4$\pm$0.2 & 4.5$\pm$0.2   & SDSSCEJ232.866760-01.117970 & 0.129 & 5.70 \\
           &       &             &     &             &             &             &             &               & SDSSCEJ233.054077-01.151869 & 0.424 & 6.60 \\ 
J1538$+$5817 & 0.143 & 205$\pm$13  & 222 & 7.9$\pm$2.5 & 8.9$\pm$1.1 & 2.8$\pm$0.7 & 4.8$\pm$0.2 & 4.8$\pm$0.3   & None & $-$ & $-$  	\\
J1618$+$4353 & 0.199 &     $-$     & $-$ & 6.0$\pm$1.9 & 10.3$\pm$1.3 & 3.7$\pm$0.8 & 4.1$\pm$0.2 & 4.1$\pm$0.2  & None & $-$ & $-$  \\
J1621$+$3931 & 0.245 & 253$\pm$21  & 285 & 2.3$\pm$0.7 & 7.8$\pm$0.9 & 2.3$\pm$0.6 & 2.6$\pm$0.1 & 4.1$\pm$0.2   & None & $-$ & $-$  	\\
J1627$-$0053 & 0.208 & 312$\pm$16  & 274 &     $-$     &     $-$     &     $-$     &     $-$     &     $-$       & None & $-$ & $-$  	\\
J1630$+$4520 & 0.248 & 297$\pm$17  & 311 & 2.0$\pm$0.6 & 9.8$\pm$1.1 & 1.7$\pm$0.5 & 2.8$\pm$0.1 & 4.2$\pm$0.2   & None & $-$ & $-$  	\\
J1636$+$4707 & 0.228 & 250$\pm$16  & 247 & 10.0$\pm$3.2 & 7.8$\pm$1.3 & 3.9$\pm$0.8 & 3.9$\pm$0.1 & 4.0$\pm$0.3  & MaxBCGJ249.00650+47.11864 & 0.235 & 0.40 \\
J1718$+$6424 & 0.090 &     $-$     & $-$ & 10.0$\pm$3.2 & 10.5$\pm$2.0 & 4.1$\pm$0.8 & 3.6$\pm$0.2 & 4.2$\pm$0.3 & SDSS-C43010 & 0.087 & 0.70 \\
           &       &             &     &             &             &             &             &               & NSCJ171819+642403 & 0.118 & 2.10 \\
J2141$-$0001 & 0.138 & 195$\pm$15  & $-$ & 2.6$\pm$0.8 & 7.8$\pm$0.9 & 2.3$\pm$0.6 & 5.2$\pm$0.3 & 4.1$\pm$0.2   & None & $-$ & $-$  	\\
J2238$-$0754 & 0.137 & 211$\pm$12  & 238 & 1.6$\pm$0.5 & 7.0$\pm$0.6 & 1.6$\pm$0.5 & 4.1$\pm$0.2 & 4.6$\pm$0.2   & None & $-$ & $-$  	\\
J2300$+$0022 & 0.228 & 301$\pm$18  & 301 &     $-$     &     $-$     &     $-$     &     $-$     &     $-$       & None & $-$ & $-$  	\\
J2302$-$0840 & 0.090 & 253$\pm$13  & $-$ & 3.4$\pm$1.1 & 10.1$\pm$2.1 & 3.6$\pm$0.7 & 5.9$\pm$0.4 & 4.4$\pm$0.3  & MaxBCGJ345.47648-08.74353 & 0.211 & 7.40 \\
J2303$+$1422 & 0.155 & 269$\pm$17  & 290 & 4.5$\pm$1.4 & 8.7$\pm$1.0 & 3.4$\pm$0.7 & 3.7$\pm$0.2 & 4.3$\pm$0.3   & MaxBCGJ345.79257+14.36653 & 0.159 & 2.80  \\	
J2321$-$0939 & 0.082 & 260$\pm$13  & 259 & 2.0$\pm$0.6 & 15.0$\pm$2.9 & 2.0$\pm$0.6 & 2.7$\pm$0.2 & 5.7$\pm$0.6  & None & $-$ & $-$  \\
J2341$+$0000 & 0.186 & 218$\pm$14  & 262 & 9.9$\pm$3.1 & 8.4$\pm$1.2 & 5.1$\pm$0.9 & 3.1$\pm$0.2 & 4.6$\pm$0.3   & SDSSCEJ355.279785-00.000869 & 0.197 & 1.20 \\
           &       &             &     &             &             &             &             &               & SDSSCEJ355.248779+00.081260 & 0.220 & 5.50 \\
           &       &             &     &             &             &             &             &               & NSCSJ234103+000250 & 0.110 & 3.30 \\
           &       &             &     &             &             &             &             &               & ABELL2644 & 0.069 & 5.30 \\

\enddata
\label{tab:env}
\tablecomments{
Col. (1): Lens ID.  
Col. (2): Lens redshift.
Col. (3): Central stellar velocity dispersion in \kms, when available. From paper V.
Col. (4): Velocity dispersion of the best fit SIE model in \kms, when available. from Paper V.
Col. (5): Surface density measured within the tenth nearest neighbor, in Mpc$^{-2}$.
Col. (6): Average $\Sigma_{10}$ for twins, in Mpc$^{-2}$.
Col. (7): Surface density of neighbors within 1$h^{-1}$Mpc radius, in Mpc$^{-2}$.
Col. (8): Average D$_1$ for random lines of sight, in Mpc$^{-2}$. Equal to average $\Sigma_{10}$ for random lines of sight.
Col. (9): Average D$_1$ for twins, in Mpc$^{-2}$.
Col. (10): Name of the clusters known to NED, within $10'$. Catalog references: {\it SDSSCE} \citet{Got++02}; {\it ZwCl} \citet{ZHW61}; {\it MaxBCG} \citet{Koe++07}; {\it NSC} \citet{Lop++04}; {\it Abell} \citet{ACO89}; {\it SDSS-C4} \citet{Mil++05}.
Col. (11): Redshift of the known cluster. Photometric redshifts are listed when spectroscopic redshifts are not available.
Col. (12): Angular distance to the known cluster in arcminutes.}
\end{deluxetable}

\clearpage

\begin{figure}
\begin{center}
\resizebox{\columnwidth}{!}{\includegraphics{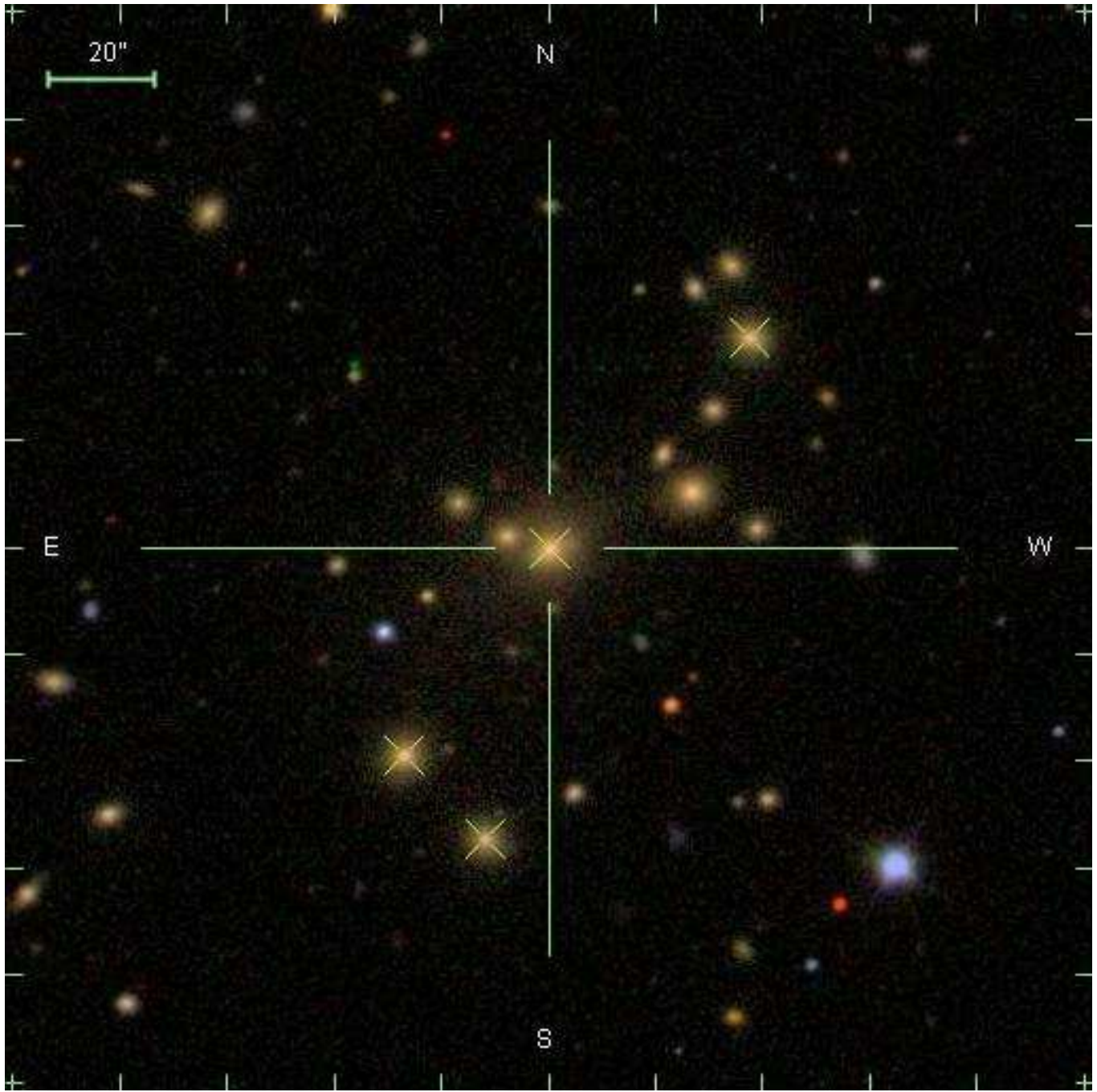}}
\end{center}
\figcaption{Field around the lens J1143-0144 as imaged by the SDSS.
Spectroscopic targets are identified by crosses. North is up, East is
left. The redshifts are consistent with that of the main lens (0.106;
at the center of the field); from left to right they are 0.104, 0.106,
0.109). \label{fig:1143}}
\end{figure}

\begin{figure}
\begin{center}
\resizebox{\columnwidth}{!}{\includegraphics{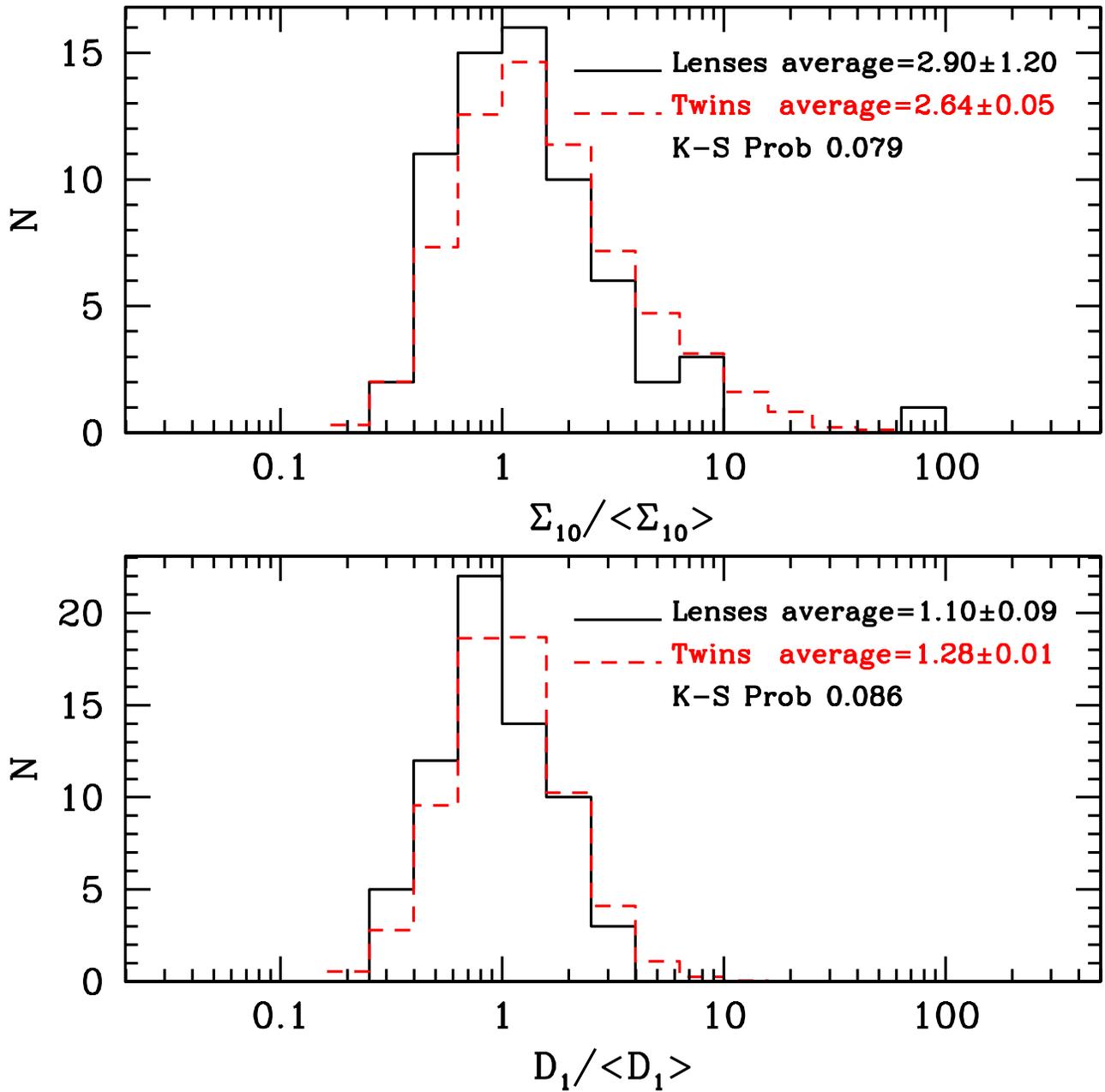}}
\end{center}
\figcaption{Distribution of overdensity of galaxies for the SLACS
sample (solid histogram) and the control ``twins'' (dashed
histogram). Two measures of environment are used: 1) $\Sigma_{10}$ the
number density within the area enclosing the ten nearest neighbors
(upper panel); 2) D$_1$ the number density of galaxies inside a cone
of radius one $h^{-1}$ Mpc. Both quantities are normalized to the
average values for the universe using random lines of sight. Both
lenses and twins are found in overdense regions as expected for
massive early-type galaxies. However, the two distributions are
statistically indistinguishable, as discussed in Section~3. Within the
uncertainties, lenses live in the same environments as non-lens
early-type galaxies.
\label{fig:plotlogs2}}
\end{figure}

\begin{figure}
\begin{center}
\resizebox{\columnwidth}{!}{\includegraphics{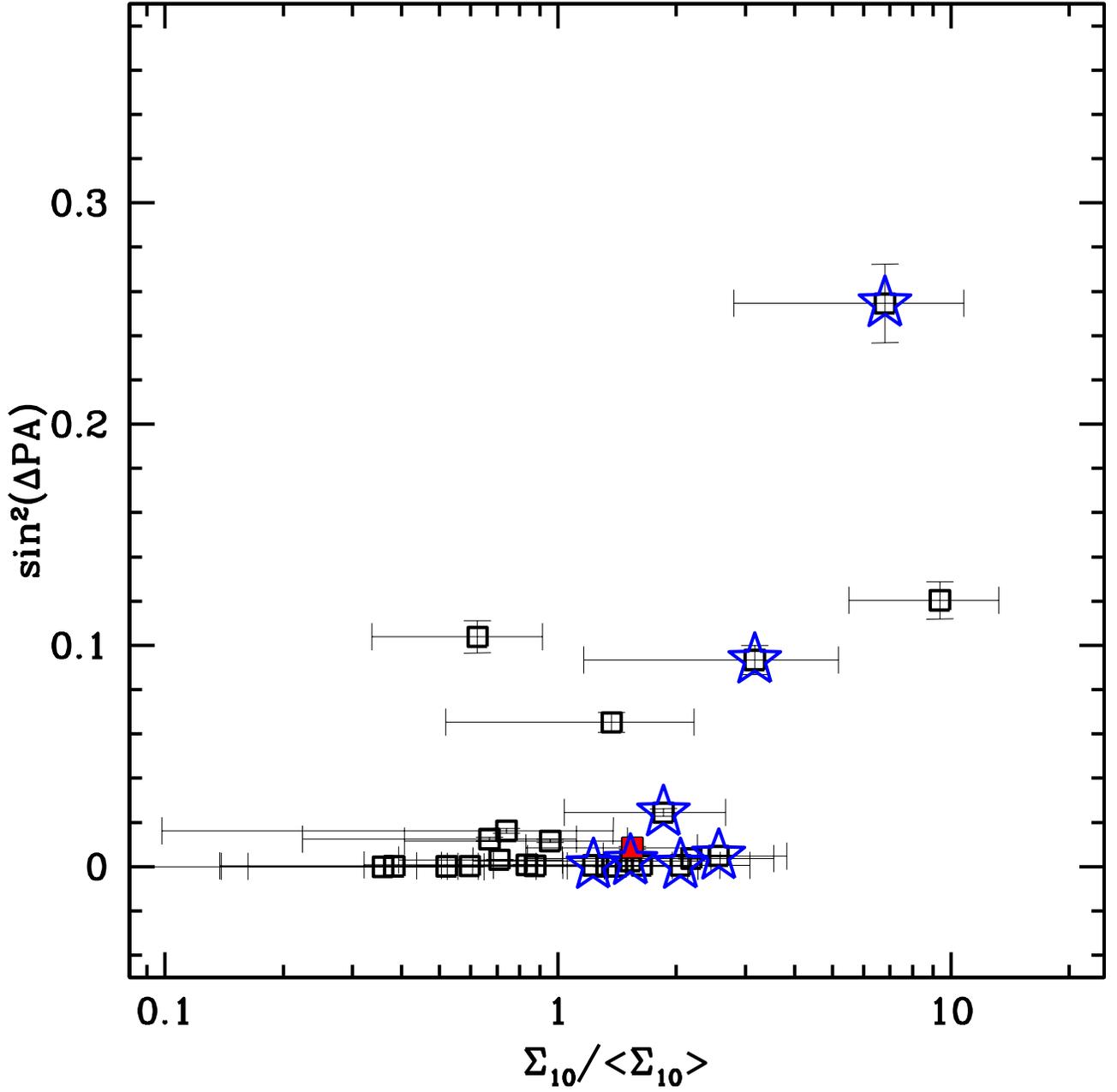}}
\end{center}
\figcaption{Misalignment between the major axis of the optical light
and that of the best fitting singular isothermal ellipsoid
($\Delta$PA) as a function of local overdensity.  The subset of
objects for which position angles can be accurately measured is shown
(i.e. axis ratio of the optical light $<0.9$ and belonging to the
``ring subsample'' as described in Paper VII). As in the rest of the
paper red solid symbols identify ``satellite'' lens galaxies while
black empty symbols identify ``central'' lens galaxies. Blue empty
stars identify lenses associated with known clusters. The misalignment
observed in overdense regions can be explained as due to randomly
aligned external shear of order $\sim0.06$.
\label{fig:S10DPA}}
\end{figure}

\begin{figure}
\begin{center}
\resizebox{\columnwidth}{!}{\includegraphics{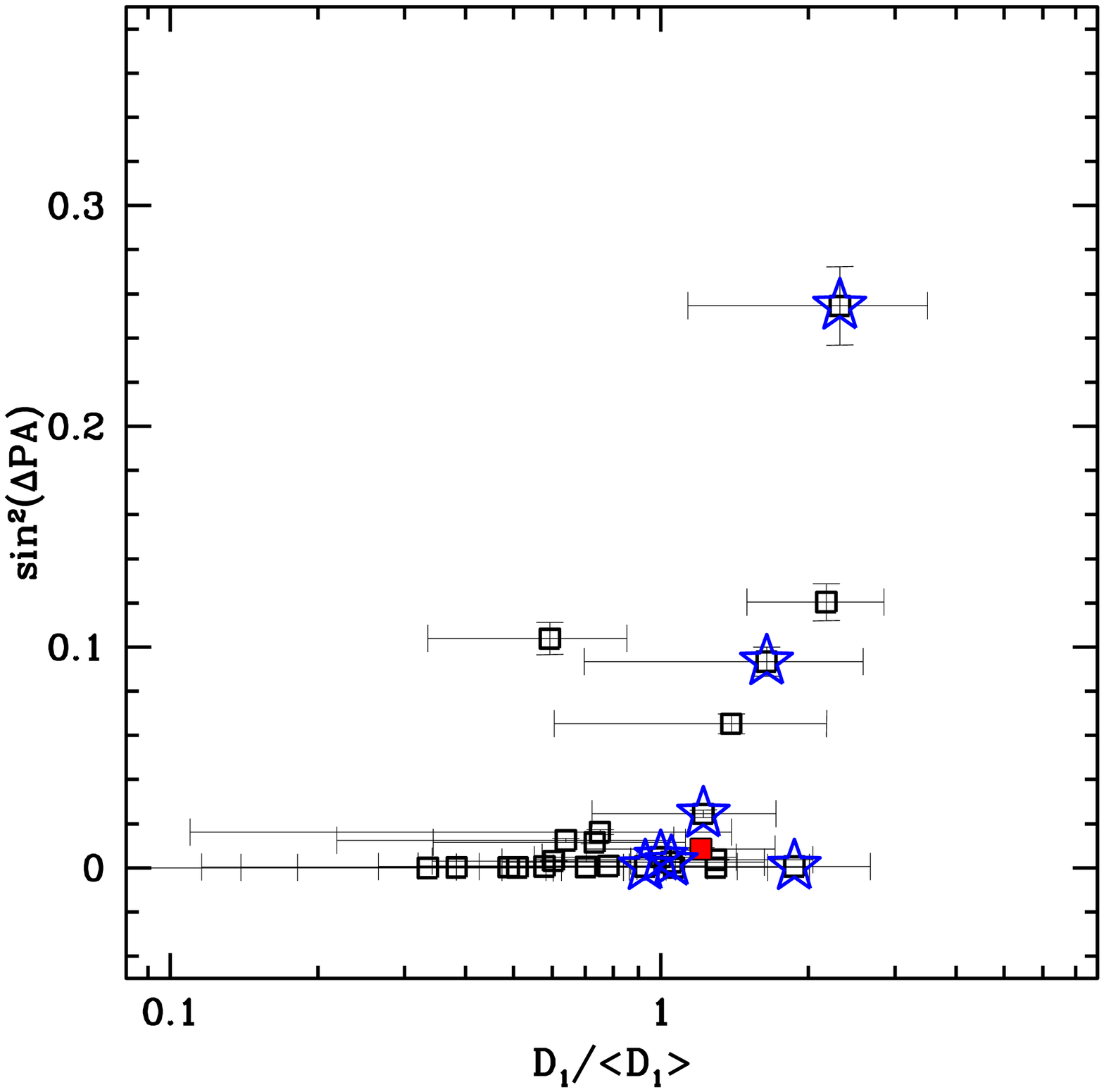}}
\end{center}
\figcaption{As in Figure~\ref{fig:S10DPA} for global overdensity
instead of local overdensity.
\label{fig:D1DPA}}
\end{figure}

\begin{figure}
\begin{center}
\resizebox{\columnwidth}{!}{\includegraphics{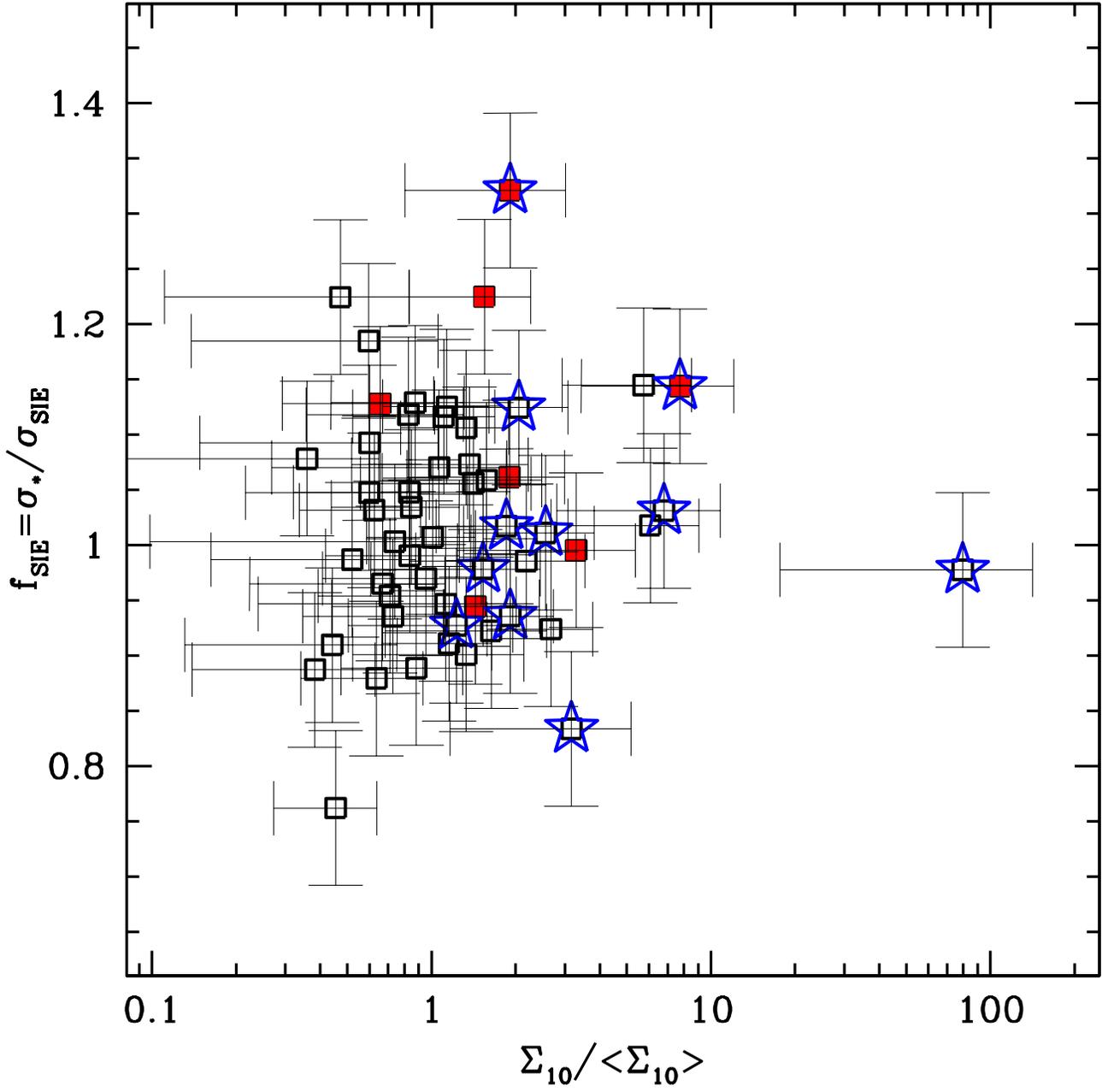}}
\end{center}
\figcaption{Ratio between central stellar velocity dispersion and that
of the best fitting singular isothermal ellipsoid as a function of the
local overdensity parameter. No significant correlation is
found. ``Satellite'' lenses are indicated by red solid symbols, while
black empty symbols indicate ``central'' galaxies.
\label{fig:S10ff1}}
\end{figure}

\begin{figure}
\begin{center}
\resizebox{\columnwidth}{!}{\includegraphics{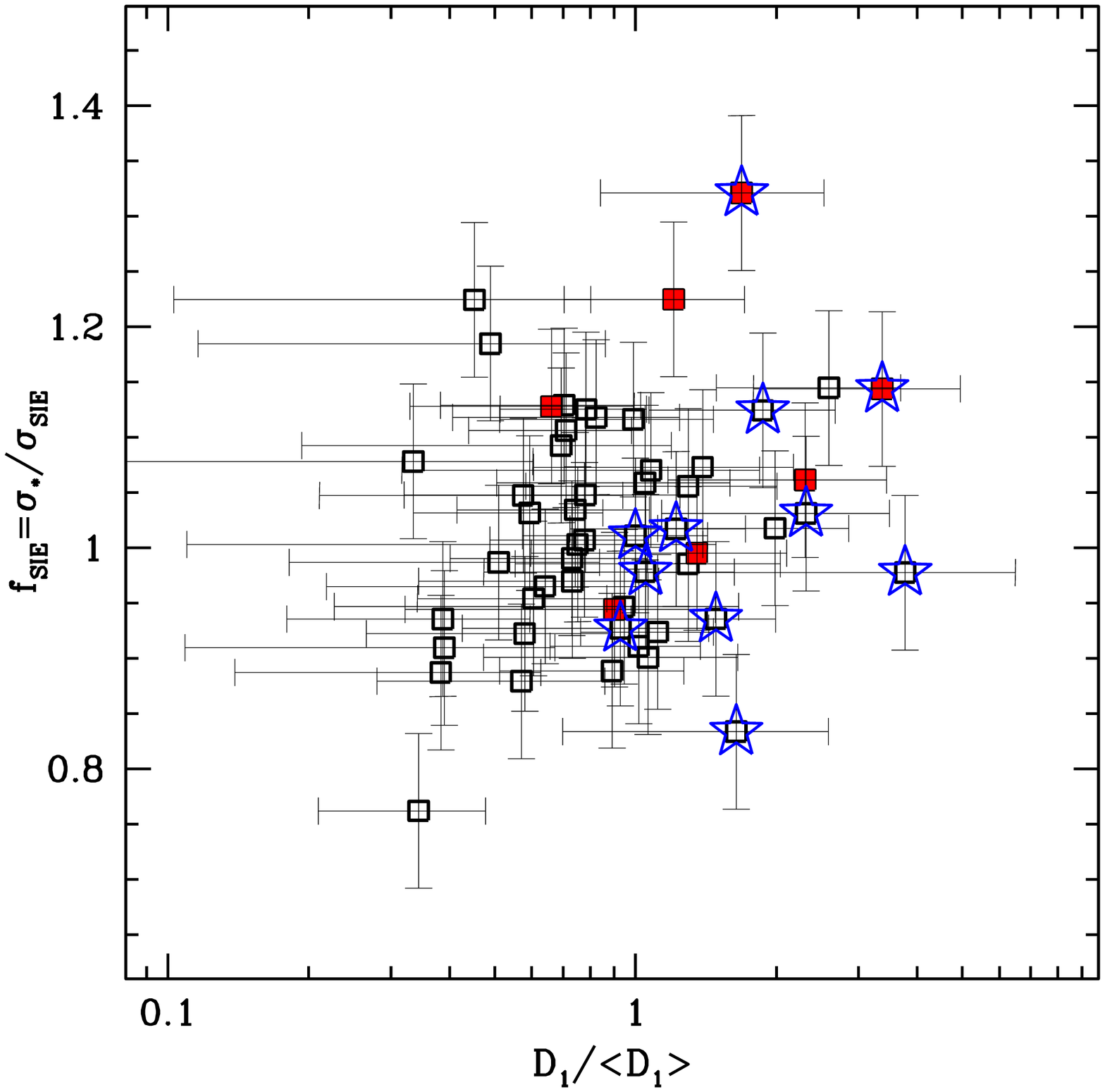}}
\end{center}
\figcaption{As in Figure~\ref{fig:S10ff1} for global overdensity
instead of local overdensity.
\label{fig:D1ff1}}
\end{figure}

\begin{figure}
\begin{center}
\resizebox{\columnwidth}{!}{\includegraphics{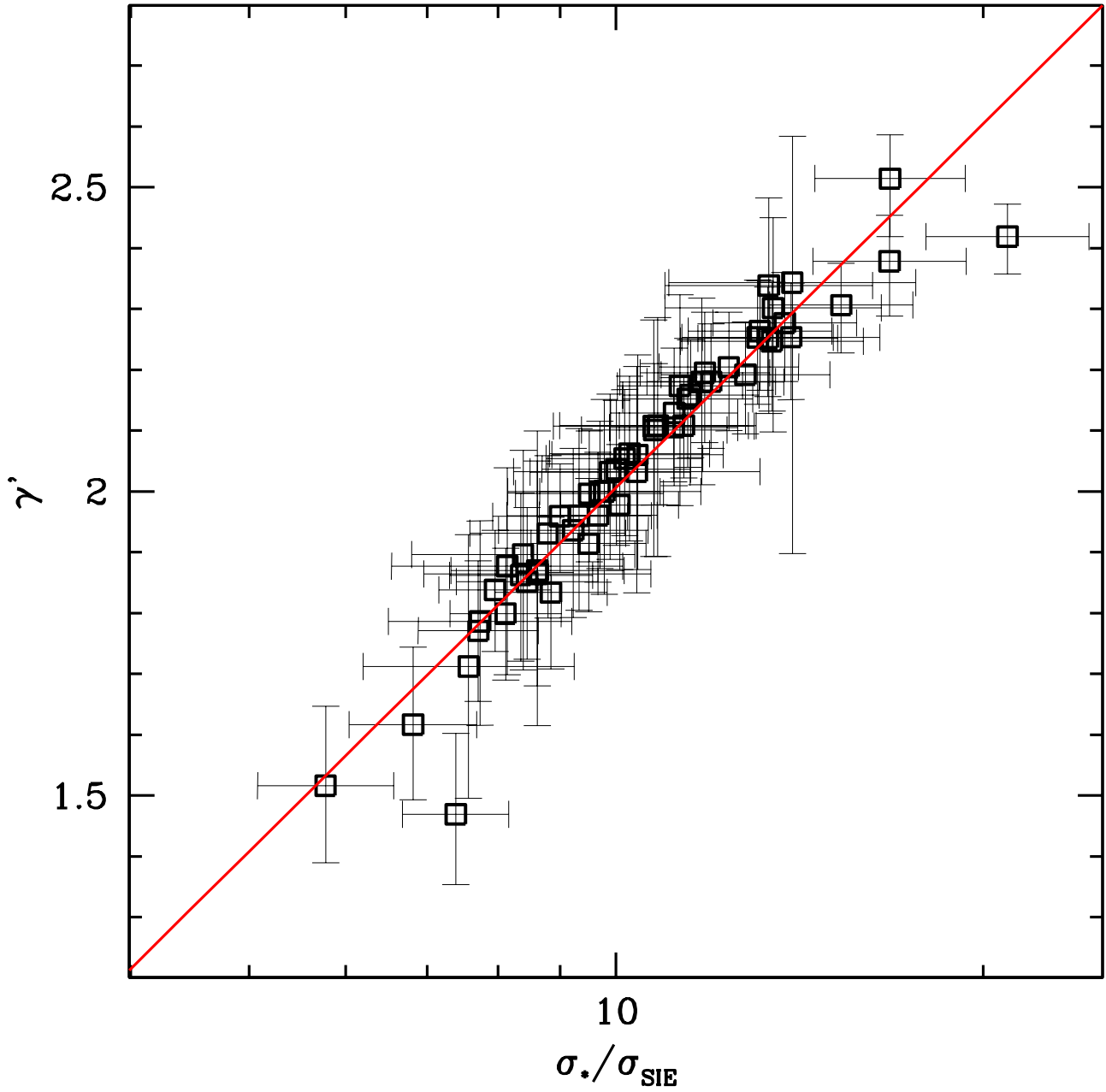}}
\end{center}
\figcaption{Transformation from $f_{\rm SIE}$ to slope of the total
mass density profile $\gamma'$ (from Paper IX, Koopmans et al.\
2008). The best fit linear relation is shown as a solid red line:
$\gamma'-2=(1.99\pm0.07)(f_{\rm SIE}-1)+(0.006\pm0.008)$.
\label{fig:fgamma}}
\end{figure}

\begin{figure}
\begin{center}
\resizebox{\columnwidth}{!}{\includegraphics{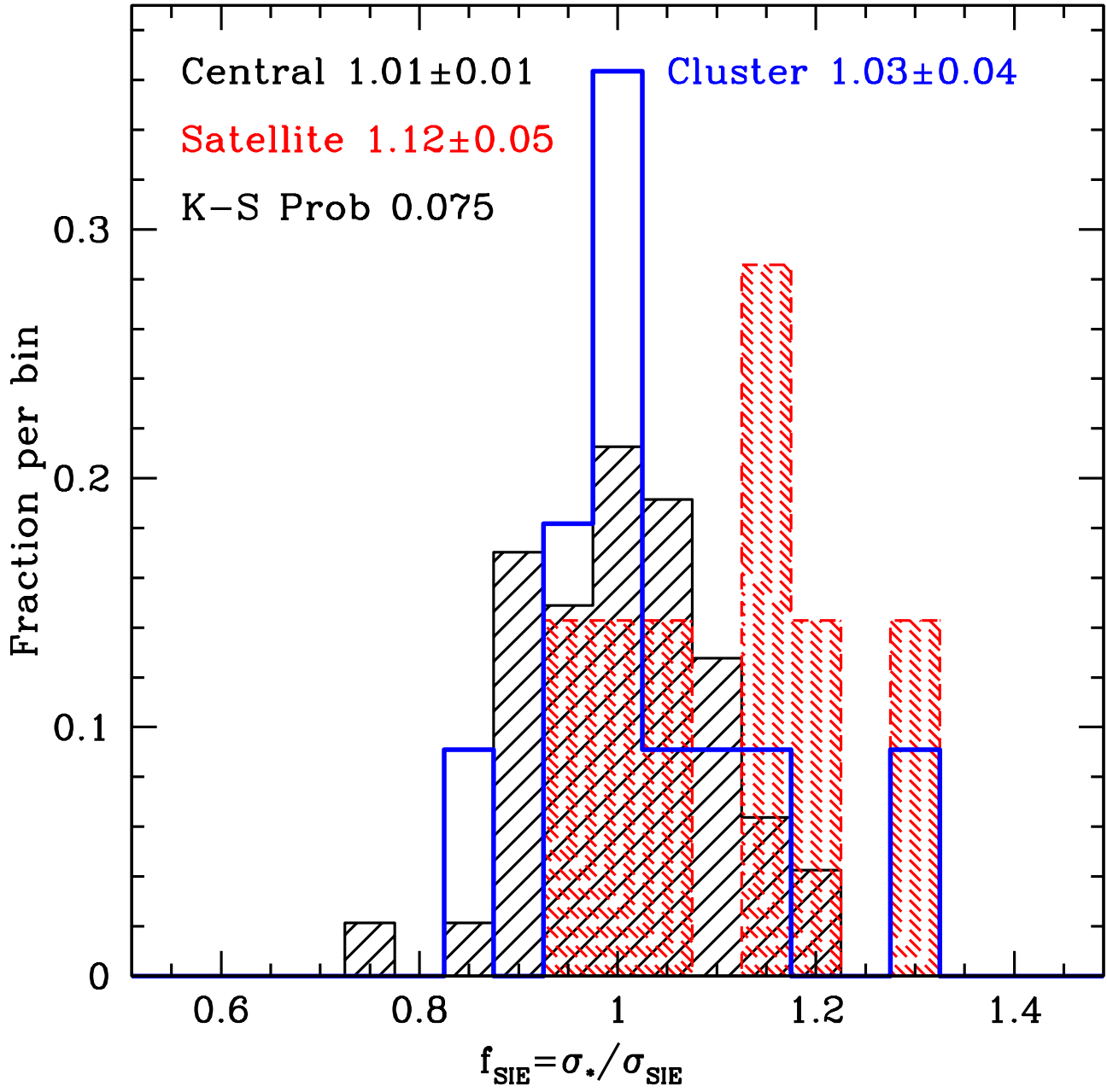}}
\end{center}
\figcaption{Distribution of the ratio between central stellar velocity
dispersion and that of the best fitting singular isothermal ellipsoid
for central (solid black histogram) and satellite (dashed red
histogram) lens galaxies. Satellite lenses have somewhat steeper
average density slopes. See \S~\ref{ssec:ff} for discussion.
\label{fig:histonbobjff1}}
\end{figure}

\begin{figure}
\begin{center}
\resizebox{\columnwidth}{!}{\includegraphics{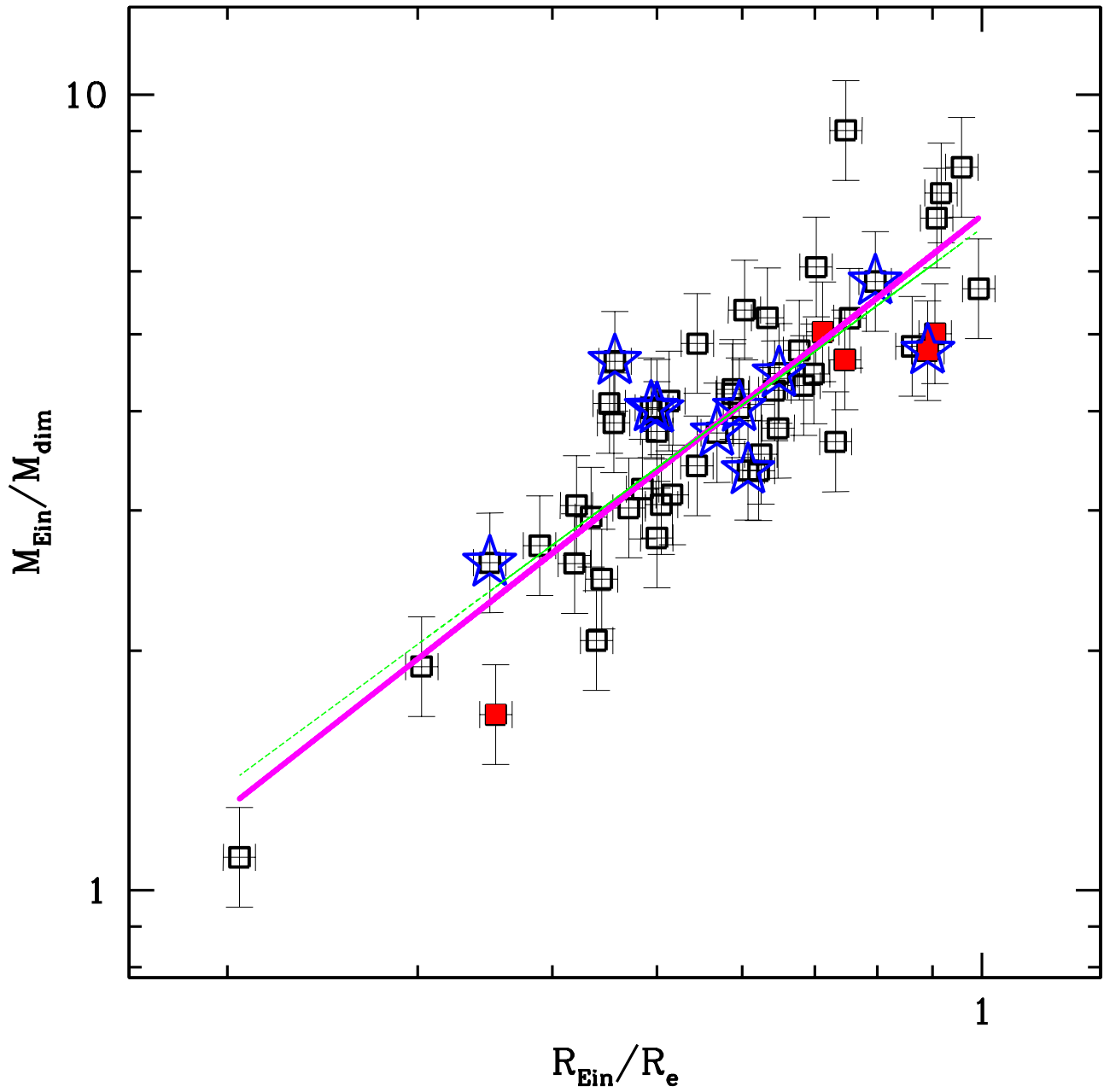}}
\end{center}
\figcaption{Projected mass inside the Einstein radius, normalized by
dimensional mass, as a function of the Einstein radius in units of the
effective radius. For an isothermal model a linear relation is
expected (thin dotted green line; Bolton et al.\ 2008b). The best fit
line (thick solid magenta line) is consistent with a linear
relation. Symbols, as in the other figures, identify satellite
galaxies (red filled squares) and galaxies associated with known
clusters (blue open stars).
\label{fig:M1}}
\end{figure}

\begin{figure}
\begin{center}
\resizebox{\columnwidth}{!}{\includegraphics{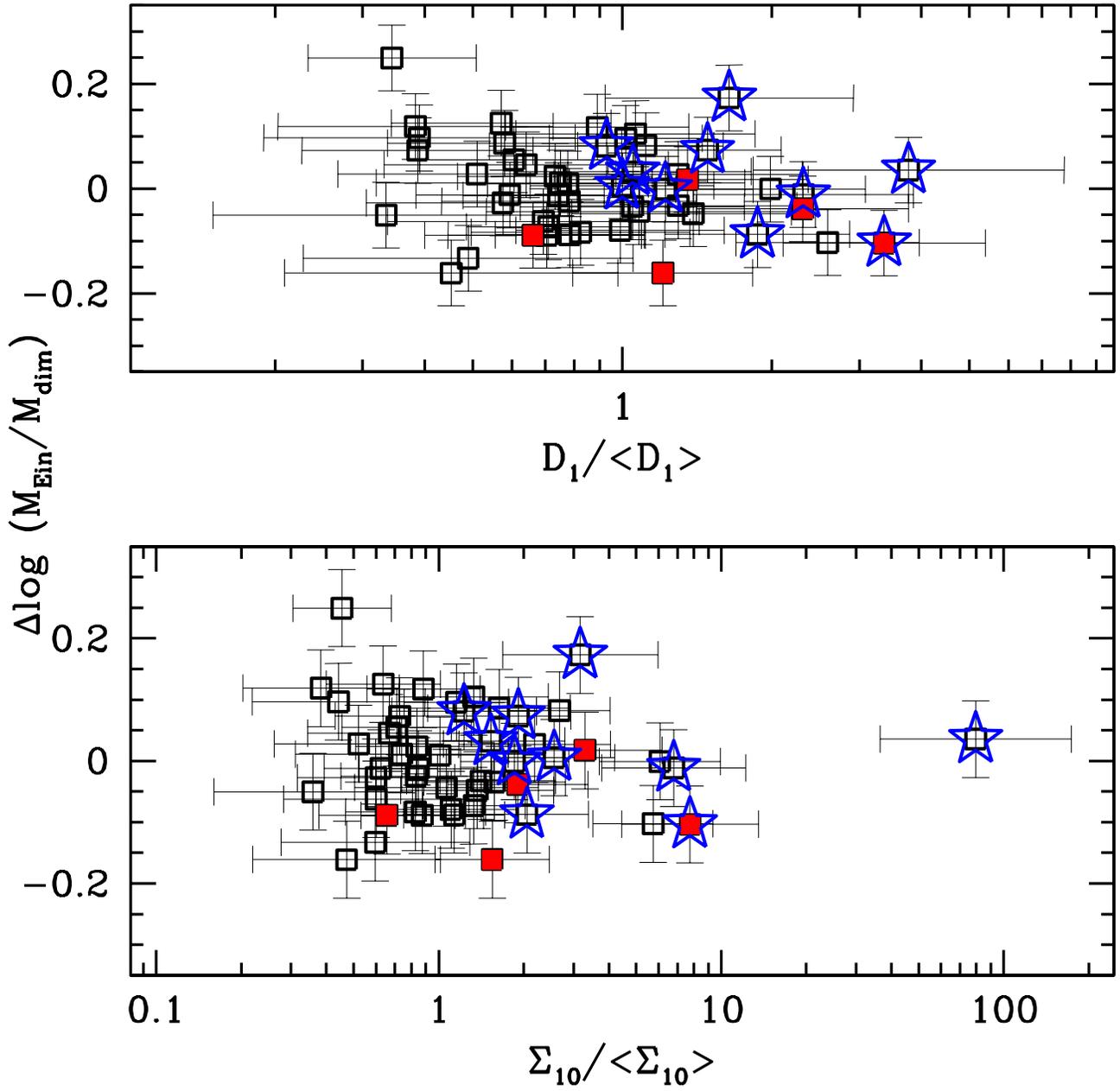}}
\end{center}
\figcaption{Residuals from the isothermal scaling ($\Delta \log M_{\rm
Ein }/M_{\rm dim}$) as a function of local (bottom) and global (top)
overdensity. Symbols, as in the other figures, identify satellite
galaxies (red filled squares) and galaxies associated with known
clusters (blue open stars).
\label{fig:M3}}
\end{figure}

\end{document}